\documentclass[12pt]{iopart}

\usepackage{iopams}
\usepackage{graphicx}
\usepackage[utf8]{inputenc}
\usepackage[colorlinks=true, pdfstartview=FitV, linkcolor=blue, citecolor=blue]{hyperref}
\usepackage{color}
\usepackage{epstopdf}
\usepackage{array}
\begin{document}

\title[Electronic and Phonon Instabilities in Bilayer Graphene under Applied External Bias]{Electronic and Phonon Instabilities in Bilayer Graphene under Applied External Bias}

\author{E. Lora da Silva$^{1,2}$, M. C. Santos$^3$, J. M. Skelton$^4$, Tao Yang$^{5,6}$, T. Santos$^7$, S. C. Parker$^2$, A. Walsh$^8$}

\address{$^1$Instituto de Diseño para la Fabricación y Producción Automatizada, and MALTA Consolider Team, Universitat Politècnica de València, 46022 València, Spain}
\address{$^2$Department of Chemistry, University of Bath, Bath BA2 7AY, United Kingdom}

\address{$^3$Department of Physics, University of Coimbra, 3004-516 Coimbra, Portugal}
\address{$^4$School of Chemistry, University of Manchester, Oxford Road, M13 9PL, United Kingdom}
\address{$^5$Key Laboratory of Biofuels, Qingdao Institute of Bioenergy and Bioprocess Technology, Chinese Academy of Sciences, Qingdao, 26610, China}
\address{$^6$TEMA-NRG, Department of Mechanical Engineering, University of Aveiro, 3810-193 Aveiro, Portugal }
\address{$^7$CICECO, Department of Physics, University of Aveiro, 3810-193 Aveiro, Portugal}
\address{$^8$Department of Materials, Imperial College London, London SW7 2AZ, United Kingdom}

\ead{elds22@bath.ac.uk}

\vspace{10pt}
\begin{indented}
\item[]September 2018
\end{indented}

\begin{abstract}
We have performed electronic-structure and lattice-dynamics calculations on the AB and AA structures of bilayer graphene.
We study the effect of external electric fields and compare results obtained with different levels of theory to existing theoretical and experimental results.
Application of an external field to the AB bilayer alters the electronic spectrum, with the bands changing under bias from a parabolic to a "Mexican hat" structure.
This results in a semi-metal-to-semiconductor phase transition, with the size of the induced electronic band-gap being tuneable through the field strength.
A reduction of continuous symmetry from a hexagonal to a triangular lattice is also evidenced through in-plane electronic charge inhomogeneities between the sublattices.
When spin-orbit coupling is turned on for the AB system, we find that the bulk gap decreases, gradually increasing for larger intensities of the bias.
Under large bias the energy dispersion recovers the Mexican hat structure, since the energy interaction between the layers balances the coupling interaction.
We find that external bias perturbs the harmonic phonon spectra and leads to anomalous behaviour of the out-of-plane flexural ZA and layer-breathing ZO' modes.
For the AA system, the electronic and phonon dispersions both remain stable under bias, but the phonon spectrum exhibits zone-center imaginary modes due to layer-sliding dynamical instabilities.
\end{abstract}

% \vspace{2pc}


%\maketitle


% If the command \eqnobysec is included in the preamble,equation numbering by section is obtained, e.g. (2.1), (2.2), etc
\eqnobysec

\section{Introduction}
\label{sec1}

% JMS: Apologies if I mangled the meaning, but I didn't fully understand the last sentence in the paragraph below.

\indent\indent
Among the numerous derivatives of the monolayer graphene (MLG) system, special interest has been given to the multi-layer allotropes \cite{berger04_short}, in particular \textit{Bernal} bilayer graphene with AB stacking (AB-BLG) \cite{pogorelov-prb-92-2015}.
Like ML graphene, BL graphene also displays unconventional properties \cite{Novoselov2006_bilayer_short} that are relevant to technological developments including tunnelling field-effect transistors \cite{ginaluca}, high-rate lithium-sulphur batteries \cite{kuhne, zhao.natcomm.5.3410}, nanophotonics \cite{yan_nanophotonics_2015}, sensor modelling \cite{akbari_jnano_2014}, among others.
These properties originate from the weak coupling between layers, which allows for the properties of the base ML graphene material to be retained.
Despite the similarities between ML and BL graphene, there are also significant differences between the two allotropes.
ML graphene shows a linear band dispersion near the Fermi energy, and the valence and conduction bands touch at the $K$-point (the Dirac point), yielding the characteristic dispersion of relativistic massless Dirac electrons \cite{Novoselov2005.Nature, PhysRep.648.2016}.
For unbiased AB-BLG, on the other hand, the interlayer coupling produces a parabolic-like band structure around the $K$-point. These different features result in a vanishing of the density-of-states (DOS) at the Fermi energy for the MLG \cite{PhysRep.648.2016}, in contrast to a finite DOS evidenced in the AB-BLG.

Another characteristic feature of AB-BLG is the behaviour of the system when an electric field is applied normal to the layers.
It has recently been shown that biased AB-BLG can form a Wigner crystal, due to the existence of different kinetic-energy dispersions at different electron densities \cite{PhysRevB.95.075438}.
The energy band gap can be tuned in proportion to the intensity of the applied bias \cite{pogorelov-prb-92-2015}, and two distinct zero-temperature quantum phases at different electron densities can be formed \cite{PhysRevB.95.075438, Jaeger1998}.

For the AB-BLG system, the presence of significant SOC has been evidenced by topological-insulator behaviour with a finite spin Hall conductivity \cite{0034-4885-76-5-056503}. Moreover, it has also been shown that biased BLG may exhibit two topologically-distinct phases depending on the intensity of the Rashba spin-orbit coupling (RSOC) \cite{PhysRevLett.107.256801}.
For weak coupling, the system exhibits a quantum-valley Hall state, which can then transition to a topological insulator in the presence of strong coupling effects.
It is possible to transition between these two phases by tuning the applied electric field \cite{PhysRevLett.107.256801}.
In the presence of strong RSOC, and for sufficiently short-range electron-electron interactions, the system minimises its energy by adopting broken-symmetry states (mostly those which break rotational symmetry) in the limit of low densities \cite{PhysRevB.85.035116}. These instabilities occur due to the energy dispersion having a minimum in a region of momentum-space which is bounded by two concentric circles with finite radius (annuli) \cite{PhysRevB.91.155423}.
Moreover, distortions to the Fermi surface, resulting from a momentum-space change in the Fermi radius (a Pomeranchuk instability) can reduce the lattice symmetry and lead to spontaneous longitudinal currents \cite{PhysRevB.91.155423}.

Another stacking arrangement of BL graphene, which coexists with the AB stacking, is the AA structure where the carbon atoms are positioned directly above each other in consecutive layers.
The electronic properties differ from those of AB-BLG due to the the stacking arrangement.
AA stacking has been experimentally observed in disordered or pregraphitic carbon, also known as turbostratic graphite, and can be distinguished from ML graphene by so-called tilting experiments \cite{PhysRevB.46.4531,PhysRevLett.102.015501}.
However, as the space groups of AA-BLG and MLG are the same ($P6/mmm$), similarities between the two are difficult to predict.

Between the two stacking environments, the AB stacking is the most energetically-favourable form, and is separated from the AA stacking by a small energy barrier.
Despite its instability, AA-BLG has started to receive significant attention.
The AA configuration shows unusual electronic properties, with two degenerate electronic and hole bands crossing at the Fermi energy \cite{PhysRevLett.109.206801}.
This electronic structure supports several electron and electron-phonon instabilities, which include, among others, a shear-shift instability \cite{PhysRevLett.109.206801}.
It has further been observed that small perturbations can destabilize the degenerate spectrum and generate an excitonic gap \cite{PhysRevLett.109.206801,PhysRevB.90.155415}.
.

While the AB-BLG system is well studied both experimentally and theoretically, comparatively less attention has been given to the AA stacking.
In the present work, we aim to provide more insight into the electronic and vibrational properties of biased AA-BLG, and to make a comparison to the AB-system, by employing first-principles simulation techniques.

We find that while the AB system presents variations on the electronic densities as a function of the applied bias, we observe that the AA system remains unaltered when an electric field is applied. SOC effects are also considerable for the biased AB-system, with the band-gap presenting different scaling behaviours according to the field intensities. The phonon dispersions of the biased AB system shows instabilities of the out-of-plane acoustic and optic modes, when compared to the stability of these modes for the unbiased system. On the other hand, phonon dispersions of the AA system remain stable under bias, but the phonon spectrum exhibits a zone-center imaginary mode resulting form the shear-mode instability.

\section{Theoretical Framework}
\label{sec:theory}

\begin{figure}
  \begin{center}
    \begin{minipage}{6cm}
      \begin{flushleft}
      a)
      \end{flushleft}
      \includegraphics[width=7cm]{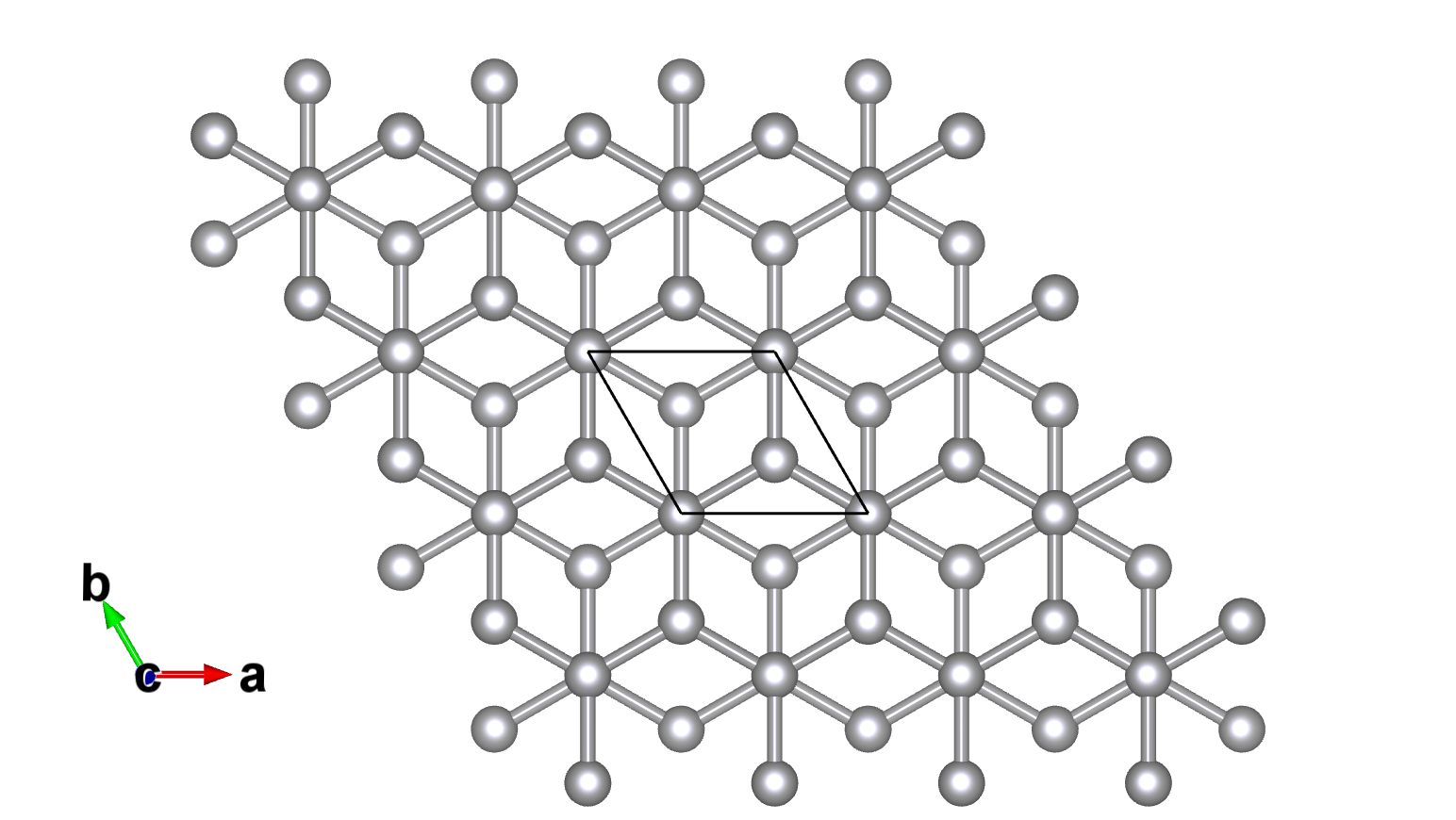}
    \end{minipage}
    \begin{minipage}{6cm}
      \begin{flushleft}
      b)
      \end{flushleft}
      \includegraphics[width=7cm]{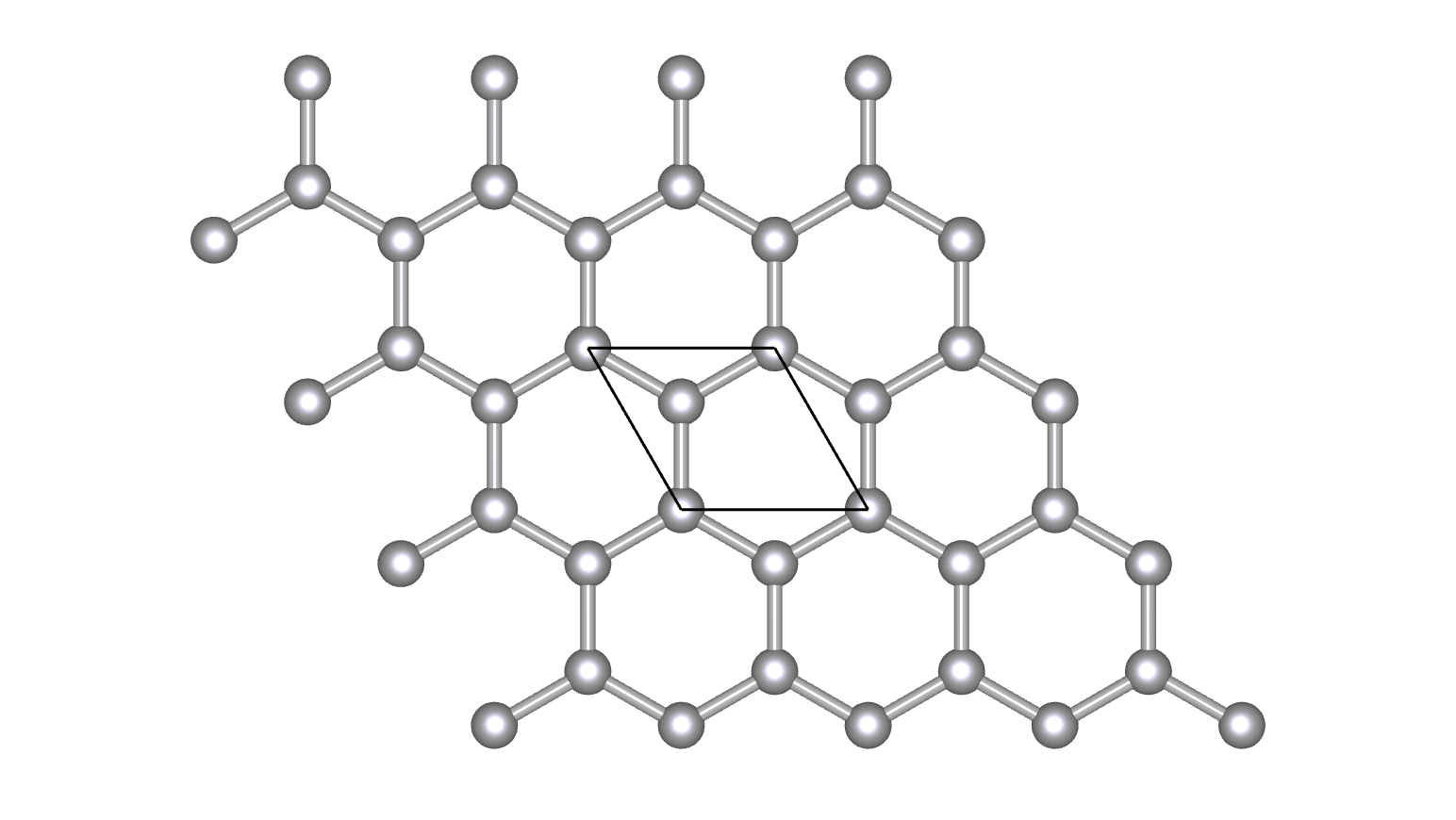}
      \end{minipage}
  \end{center}
  \caption{
    {\label{fig:BLG}}
    Supercell of the AB- (space group $P-3m1$, nº 164; a) and AA-BLG (space group $P6/mmm$, nº 191; b) systems, where the black line shows the unit-cell.
    BLG consists of two coupled monolayers of carbon atoms, each with a honeycomb crystal structure.
    In order to satisfy the translational and symmetry properties of the Bravais lattice, the honeycomb lattice can be seen as two triangular sublattices, mathematically labelled as inequivalent $A$ and $B$ lattices, each of which contains two atoms in the unit cell within each C sheet, with atom [a$_1$, a$_2$] $\in $ A and [b$_1$,b$_2$] $\in $ B for layer $1$ and $2$.
    The layers of the AB-BLG are arranged in such a way that one of the atoms from the lower-layer b$_1$ is directly below atom a$_2$ from the upper layer, and the remaining two atoms, a$_1$ and b$_2$, are shifted from each other by a vector displacement \cite{PhysRep.648.2016}.
    For the AA-BLG, the carbon atoms are aligned in the consecutive layers, directly above/below each other (a$_1$ with a$_2$ and b$_1$ with b$_2$).
    }
\end{figure}

We study the electronic structure of the two different stacking environments of the BLG system (crystal structure of AB- and AA-BLG presented in figure \ref{fig:BLG}) using density-functional theory (DFT) with the Local-Density Approximation (LDA) functional.
An external electric field is applied in the direction of the interlayer plane with variable magnitude.
Lattice dynamics are performed within the harmonic approximation, which yields phonon frequencies and the constant-volume terms in the free energy from lattice vibrations.

\subsection{Density Functional Theory}
\indent\indent
Electronic-structure calculations were performed within the pseudopotential plane-wave density-functional theory (DFT) framework, as implemented in the \textit{Vienna Ab-initio Simulation Package}  (VASP) \cite{kresse-prb-54-1996,kresse-prb-47-1993,kresse-cms-6-1996} code.
The Ceperley and Alder form of the Local-Density Approximation (LDA) functional, parametrised by Perdew and Zunger \cite{PhysRevB.23.5048}, was used in conjunction with projector augmented-wave (PAW) pseudopotentials \cite{PhysRevB.59.1758,PhysRevB.50.17953}.
We selected the LDA functional because it is known to perform well at capturing the interlayer distance in graphite and multi-layer graphene allotropes, as well as the essential physics of the electronic structure, and also performs well for calculating interatomic force constants and phonon frequencies \cite{PhysRevB.89.064305,PhysRevB.60.11427}.

A plane-wave cut-off of 800 eV was applied in all calculations; although convergence of the electronic structure was attained at a lower cut-off of $\sim$ 600 eV, a higher value was chosen to improve the description of the structural parameters and forces, which is important for accurate lattice-dynamics calculations \cite{PhysRevB.91.144107}.
The Brillouin zone (BZ) was sampled with $\Gamma$-centred Monkhorst-Pack meshes \cite{monkhorst-prb-13-1976} with 44$\times$44$\times$1 and 90$\times$90$\times$1 subdivisions for AA- and AB-BLG respectively.
It was found necessary to employ the denser \textbf{k}-point mesh for the AA-BLG model due to differences in the DFT electronic band structure relative to the spectra expected from tight-binding theory (section \ref{appendix-sec1}).
The vacuum spacing between periodic images along the $Z$ direction was set to 15 \AA~ for both configurations, and dipole corrections to the potential were applied to avoid interactions between periodic images.

Lattice-dynamics calculations were carried out using the Parlinski-Li-Kawazoe supercell finite-displacement method \cite{parlinski-prl-78-1997, chaput-prb-84-2001}, which is implemented in the Phonopy \cite{phonopy, togo-prb-78-2008} package; a detailed description of the theoretical implementation can be found in Refs. \cite{skelton-prb-89-2014, PhysRevB.91.144107}).
The interatomic force constants were obtained by performing single-point force calculations on a series of symmetry-inequivalent displaced structures and fitting the resulting force/displacement curves to a harmonic function.
VASP was used as the force calculator \cite{chaput-prb-84-2001} and the calculations were performed on 4$\times$4$\times$1 supercells using a reduced \textbf{k}-point sampling mesh of 12$\times$12$\times$1 for both phases.
For the calculations under bias, the electric field was applied during the force-constant calculations.
To construct the phonon density of states, the phonon frequencies were sampled on an interpolated 48$\times$48$\times$1 \textbf{q}-point mesh.

A non-analytical correction (NAC) was applied when computing the phonon-band dispersion \cite{yu-cardona-springer-1996} to correct for long-range Coulomb interactions.
The requisite Born effective-charges and static dielectric constant were computed using the density-functional perturbation theory (DFPT) routines implemented in VASP \cite{gajdos-prb-73-2006}.
Convergence of these quantities required increasing the \textbf{k}-point mesh for the AB system up to 80$\times$80$\times$1, whereas for the AA system the 90$\times$90$\times$1 mesh was found to be sufficient.

A bias was applied in the calculations as an external electrostatic field in the $Z$ direction and geometries were re-optimised with different intensities of the field. Born effective-charges and dielectric tensors were calculated by considering the field perturbations. For the lattice-dynamics calculations, the bias was also applied during the calculations of the force constants.

\section{Results and Discussion}
\label{sec:discuss}

\indent\indent
The lattice parameters obtained within the LDA are a$_0$=2.42 \AA~ and c$_0$=6.69 \AA~ for the AB system, and a$_0$=2.45 \AA~ and c$_0$=6.67 \AA~ for the AA system.
The intra-layer distance (C-C bond lengths) are on the order of 1.41 \AA~ in both stacking environments, and the interlayer distance was calculated to be 3.35 \AA~and 3.34 \AA~for the AB and AA configurations, respectively.
The parameters for AB-BLG are in agreement with those discussed in Ref. \cite{0957-4484-21-6-065711}, where the calculations were also performed with DFT-LDA (intra-layer distance of 1.41 \AA~ and interlayer distance of 3.31 \AA). The present interlayer parameters also compare well to experimental results, where for the Bernal graphite the value of 3.35 \AA~\cite{PhysRevB.39.12598} was observed. However, for the AA-BLG the present interlayer distance is found to be slightly lower than results found in literature: 3.59 \AA~from DFT-LDA calculations \cite{0957-4484-21-6-065711}, and 3.55 \AA~ from experimental observations on the AA graphite structure \cite{doi:10.1063/1.2975333}.

\subsection{Electronic Spectrum from a Density-Functional and Tight-Binding Perspective}

\begin{figure}
  \begin{center}
    \begin{minipage}{7.0cm}
      \begin{flushleft}
      a)
      \end{flushleft}
      \includegraphics[width=8cm]{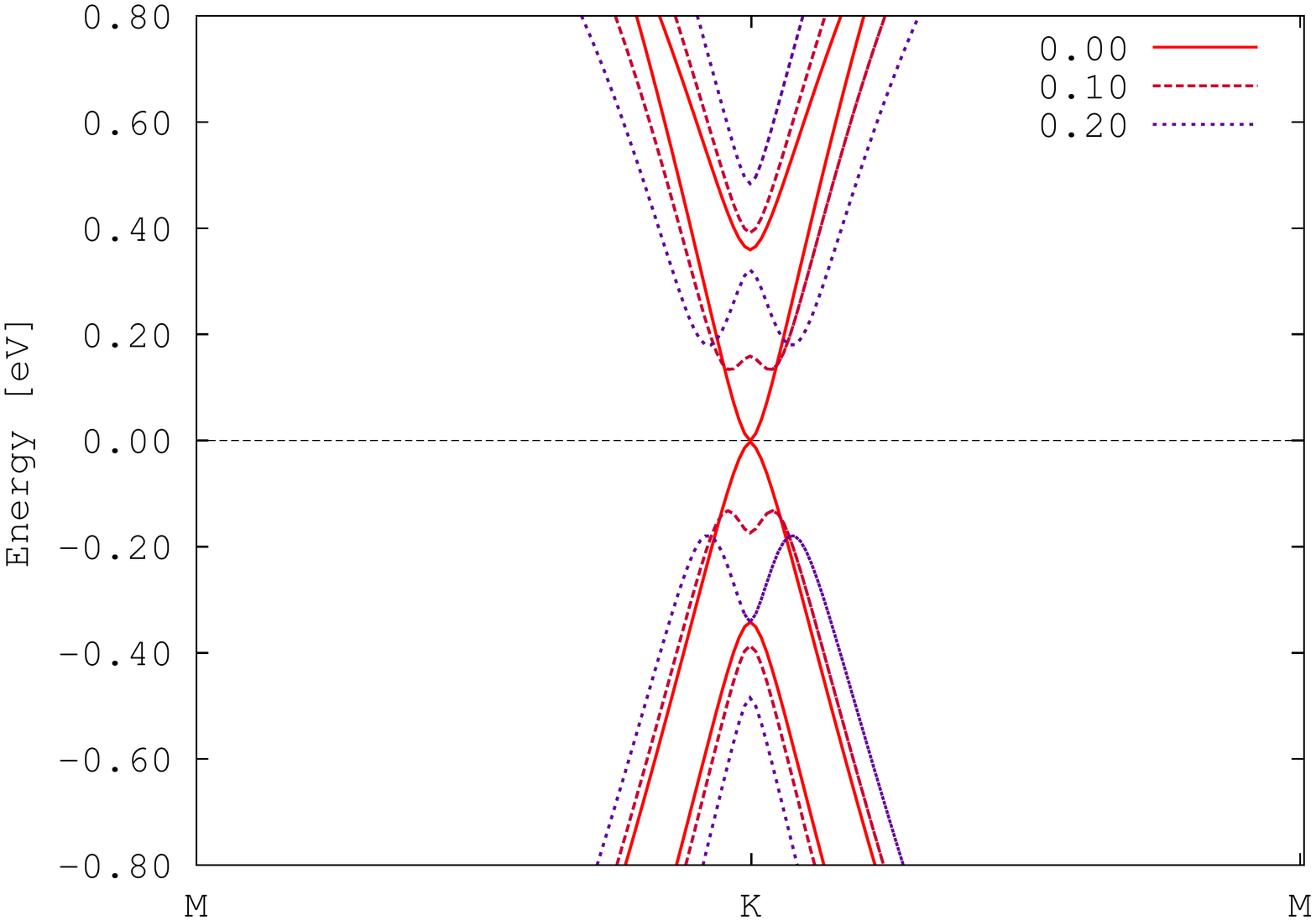}
    \end{minipage}
    \begin{minipage}{7.0cm}
      \begin{flushleft}
      b)
      \end{flushleft}
      \includegraphics[width=8cm]{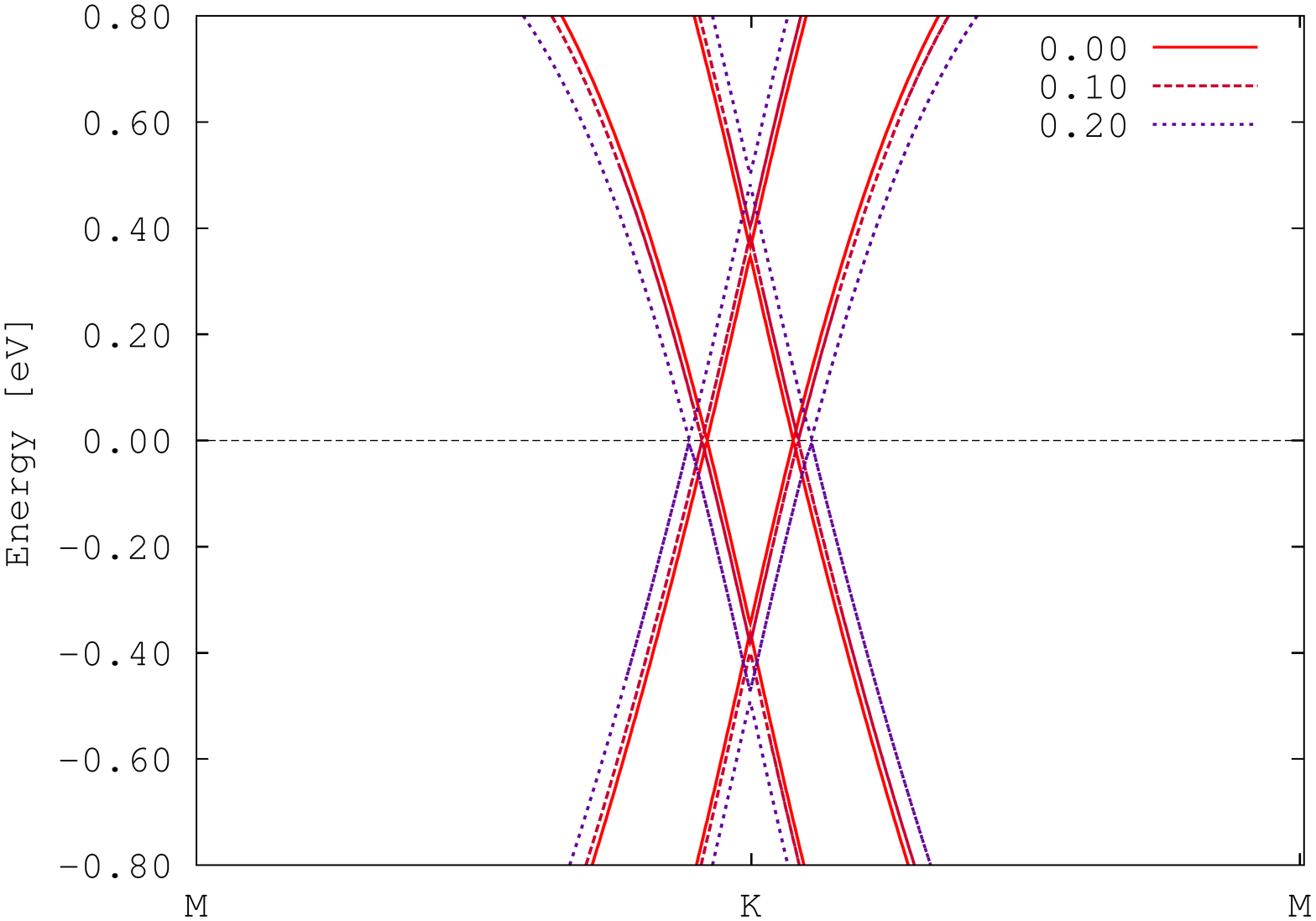}
    \end{minipage}
  \end{center}
  \caption{
  {\label{fig:elect-BS}}
  Low-energy DFT-LDA electronic band-structure of bilayer graphene with AB (a) and AA (b) stacking arrangements.
  Each dispersion is shown at different applied field intensities (label units given in eV/\AA).
  }
\end{figure}

\indent\indent
To study the electronic structure, we calculated the low-energy band dispersions using LDA-DFT with three intensities of applied electric field.
The results are presented in figure \ref{fig:elect-BS}.
For the AB-BLG configuration (Figure \ref{fig:elect-BS}.a), when $E=0$ eV/\AA, a zero-gap parabolic dispersion around the $K$-point is observed.
The LDA-DFT electronic dispersion for the AB system shows similar features to the band-structure obtained from the tight-binding Hamiltonian (figure \ref{fig:elect-TB}, Appendix A).

When a finite electric-field is applied perpendicular to the graphene layers in AB-BLG, the two layers are subject to inequivalent potentials.
This effect breaks the inversion symmetry, resulting in the opening of a single-electron gap \cite{pogorelov-prb-92-2015} at the $K$-point, which can be tuned up to mid-infrared energies ($\sim$ 300 meV) \cite{PhysRevLett.99.216802}.
A spontaneous translation symmetry breaking also occurs, resulting in a charge separation between the inequivalent sublattices with spatial in-plane charge inhomogeneities \cite{sem, PhysRevB.95.075438}.

\begin{figure}[!]
  \begin{center}
    \begin{minipage}{7cm}
      \begin{flushleft}
      a)
      \end{flushleft}
      \includegraphics[width=8.5cm]{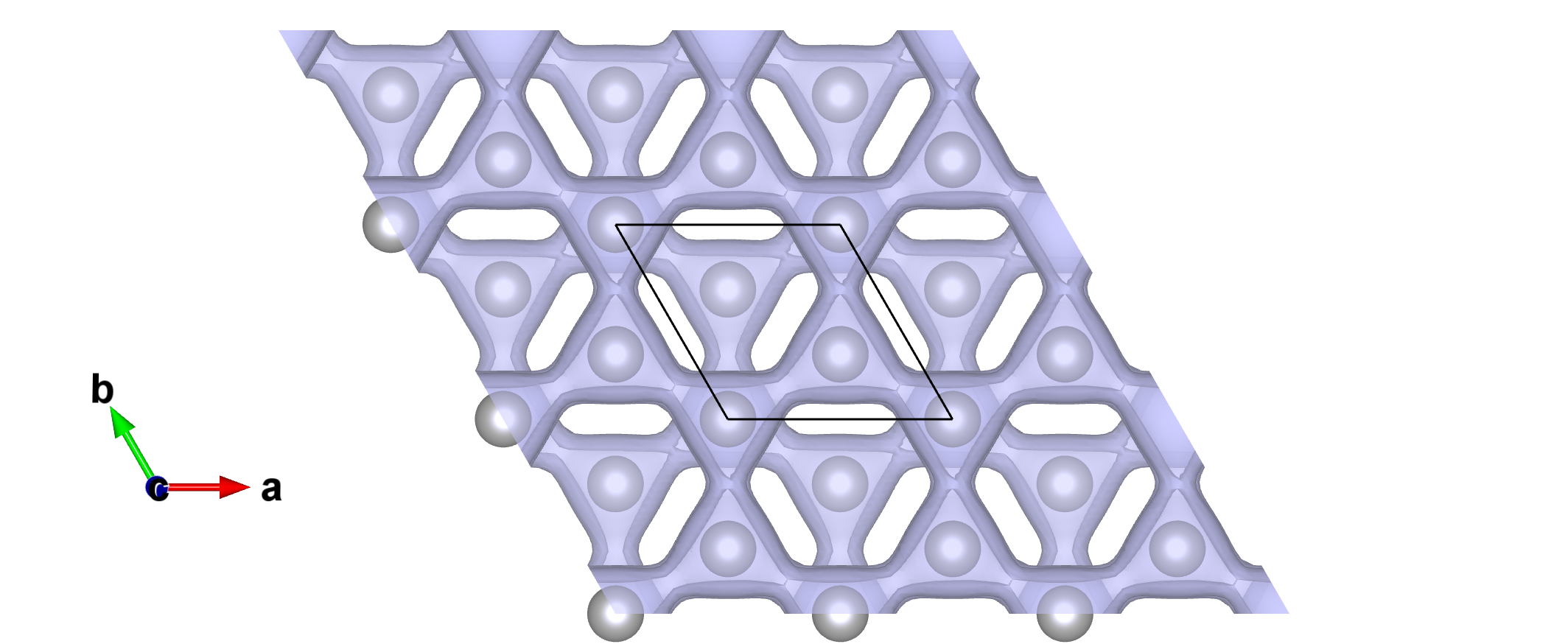}
    \end{minipage}
        \begin{minipage}{7cm}
      \includegraphics[width=8.5cm]{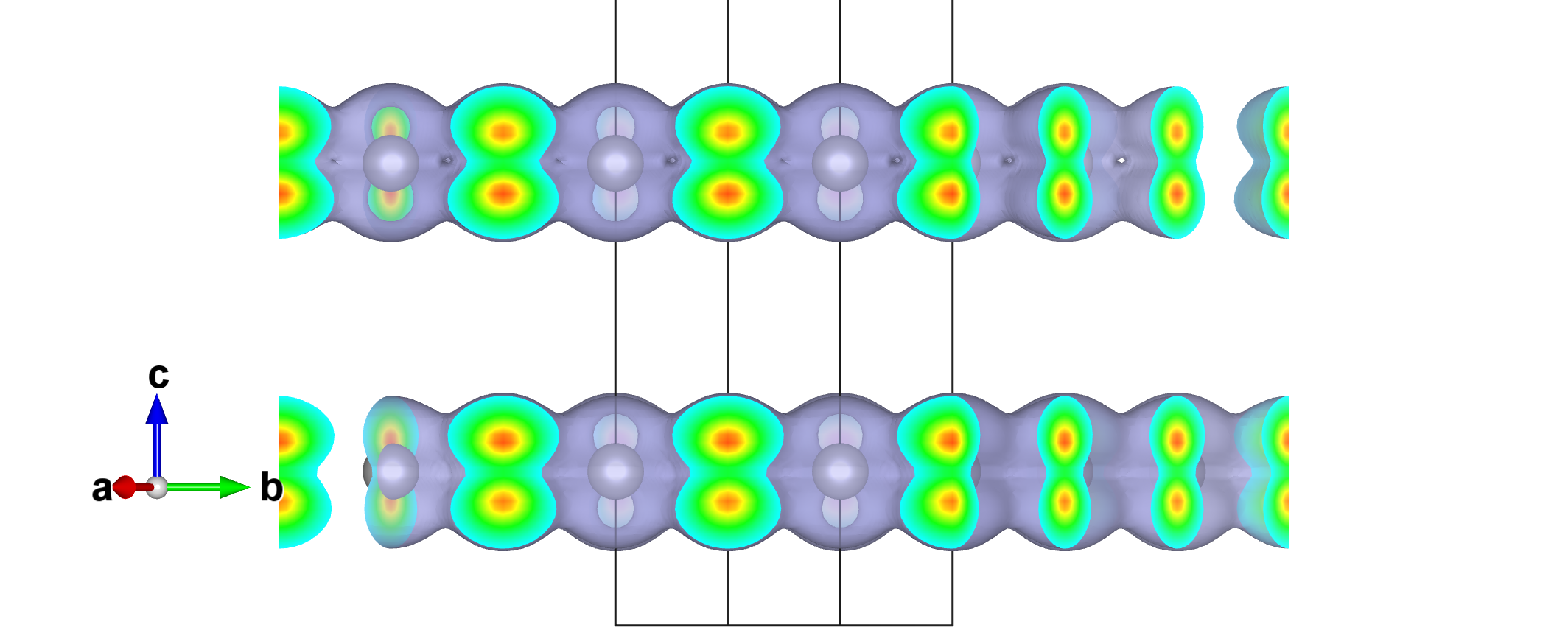}
    \end{minipage}
    \begin{minipage}{7cm}
      \begin{flushleft}
      b)
      \end{flushleft}
      \includegraphics[width=9cm]{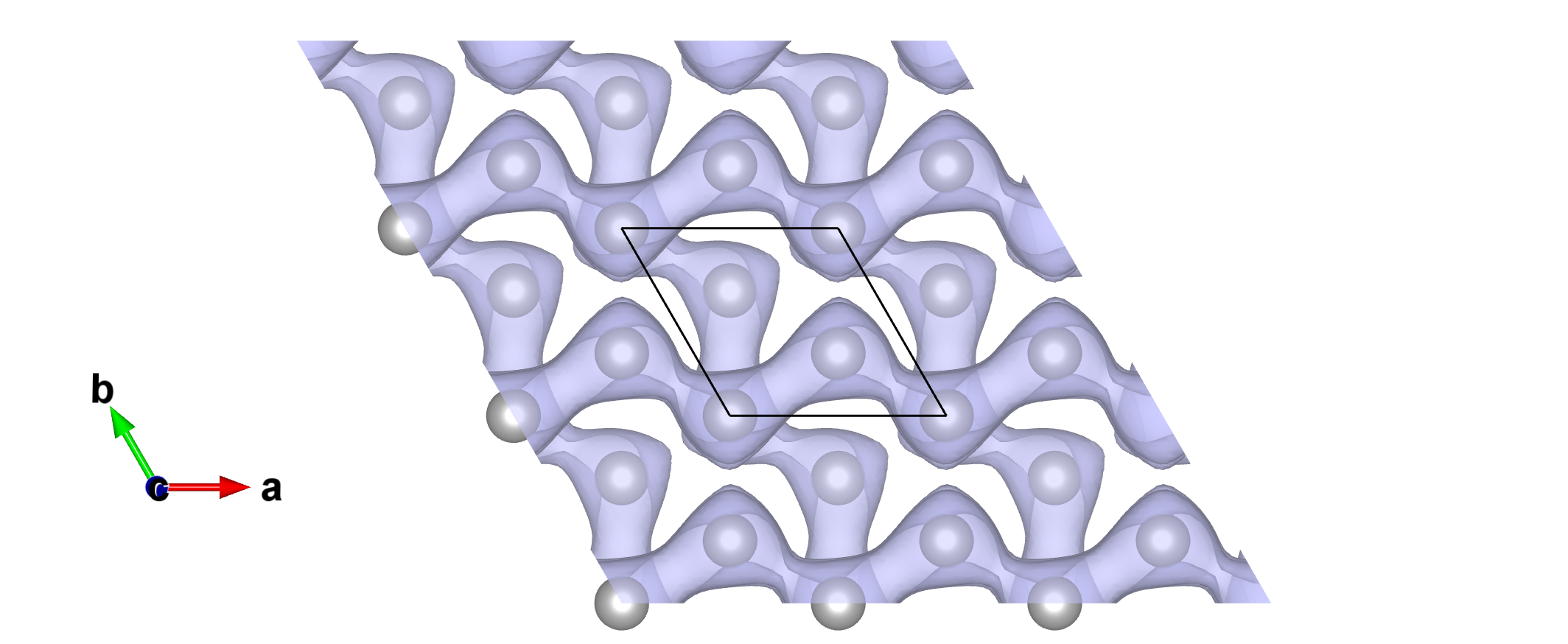}
    \end{minipage}
        \begin{minipage}{7cm}
      \includegraphics[width=9cm]{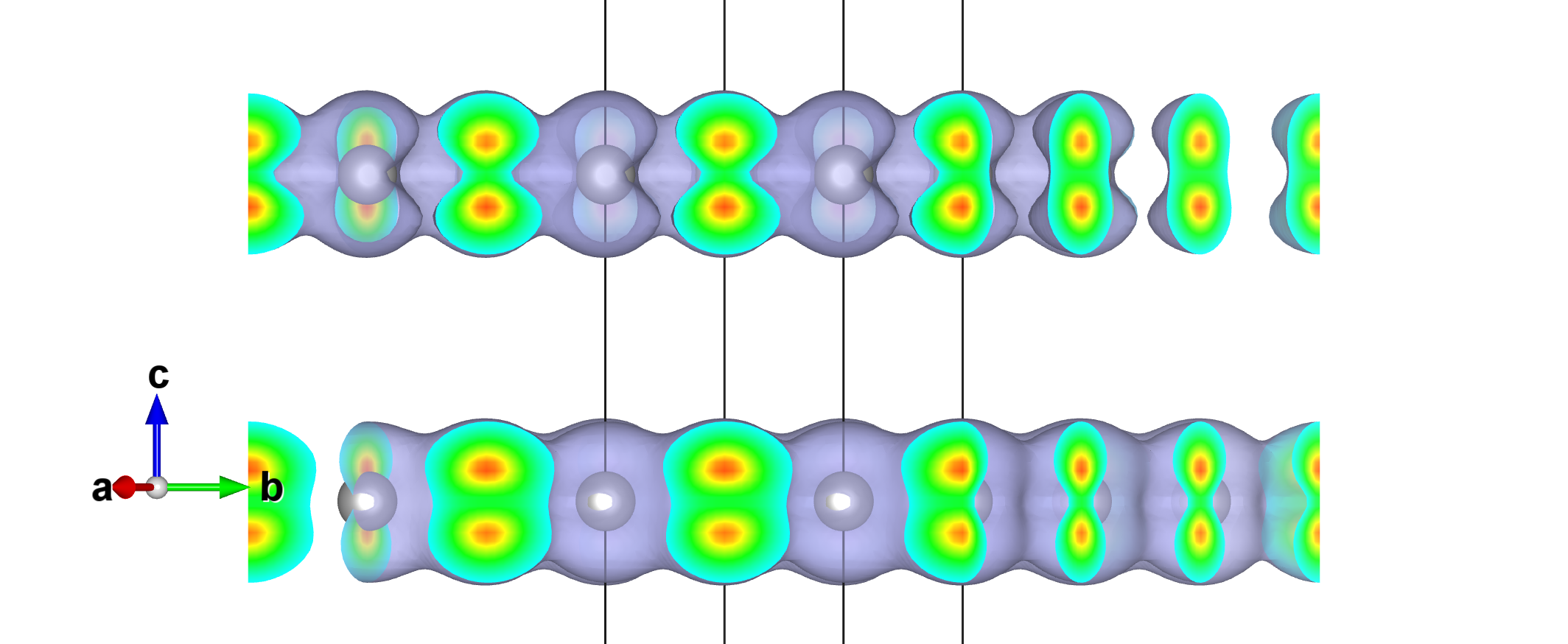}
    \end{minipage}
  \end{center}
  \caption{
    {\label{fig:isosurface}} Two different views (top and side) of the isosurfaces (value defined at 0.016) of the electron charge densities around the Fermi energy for the AB-BLG system without (a) and with (b) an applied bias (electric field intensity of 0.05 eV/\AA.}
\end{figure}

Figure \ref{fig:isosurface} plots the electron charge densities of AB-BLG in the vicinity of the Fermi energy, inspection of which reveals differences between the isosurfaces without (a) and with (b) a bias applied.
In the unbiased system (figure \ref{fig:isosurface}.a), the charge densities show hexagonal symmetry, indicating homogeneous electron delocalisation between the sublattices.
On the other hand, when an interlayer electric field is applied (figure \ref{fig:isosurface}.b), redistribution of electron densities leads to charge separation between the A and B sublattices, leading to in-plane charge inhomogeneities \cite{PhysRevB.95.075438}.% and breaking of the continuous symmetry of the triangular lattice.

The AA-stacking environment differs from the Bernal system by having a linear dispersion with two bands crossing each other at the Fermi energy \cite{PhysRep.648.2016}.
Application of an external field does not alter the width of the band gap, and electronic structure remains qualitatively the same.
This single-electron property seems to be quite stable to external bias both in the LDA calculations and also with a tight-binding Hamiltonian \cite{PhysRep.648.2016}.

These results are consistent with the electronic dispersion calculated with the tight-binding method (figure \ref{fig:elect-TB}), although, as noted above, obtaining a fully-converged dispersion from the DFT calculations required very dense \textbf{k}-point sampling.
This is because the band crossing does not occur at a high-symmetry \textbf{k}-point, and thus a dense mesh is required in order to include sufficient sampling around the feature to accurately represent the bands in the vicinity of the Fermi energy.

Under bias, the dispersion relations of the AB system show a "Mexican hat" structure \cite{pogorelov-prb-92-2015, PhysRevB.89.041405}.
With increasing field intensity, the width of the gap increases and the radius of the hat feature widens, with the two minima getting progressively further apart from the $K$-point \cite{pogorelov-prb-92-2015}.
This behaviour is consistent with the results from Ref. \cite{PhysRevB.95.075438}, which suggest that regions of the dispersion should exhibit different scaling behaviour as a function of momentum \cite{PhysRevB.95.075438}.
Moreover, controlling the magnitude of the gap through additional screening with a transverse electric field will afford control over the density of electrons \cite{PhysRevB.74.161403}.

\begin{figure}
  \begin{center}
    \begin{minipage}{7.0cm}
      \begin{flushleft}
      a)
      \end{flushleft}
      \includegraphics[width=8cm]{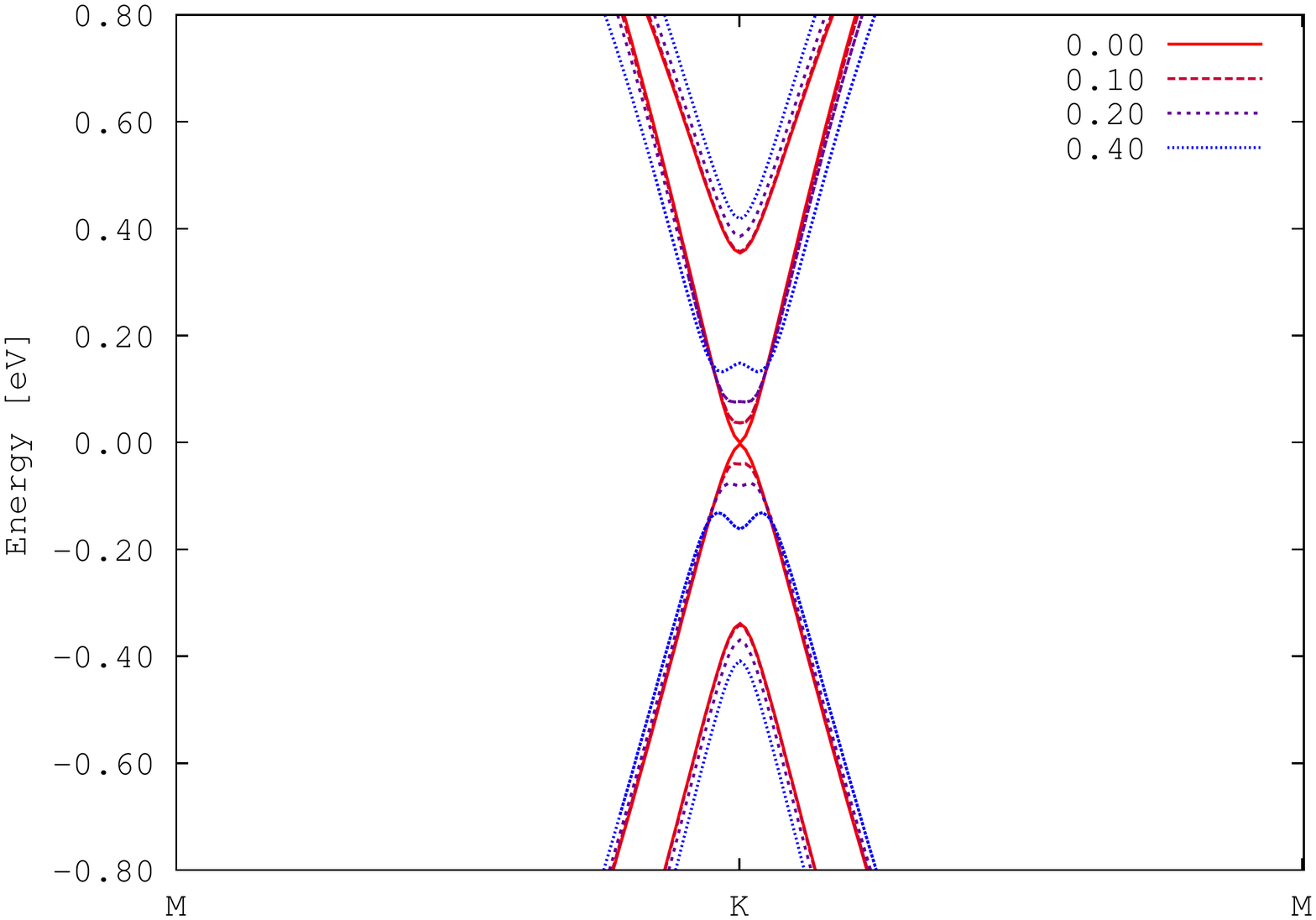}
    \end{minipage}
    \begin{minipage}{7.0cm}
      \begin{flushleft}
      b)
      \end{flushleft}
      \includegraphics[width=8cm]{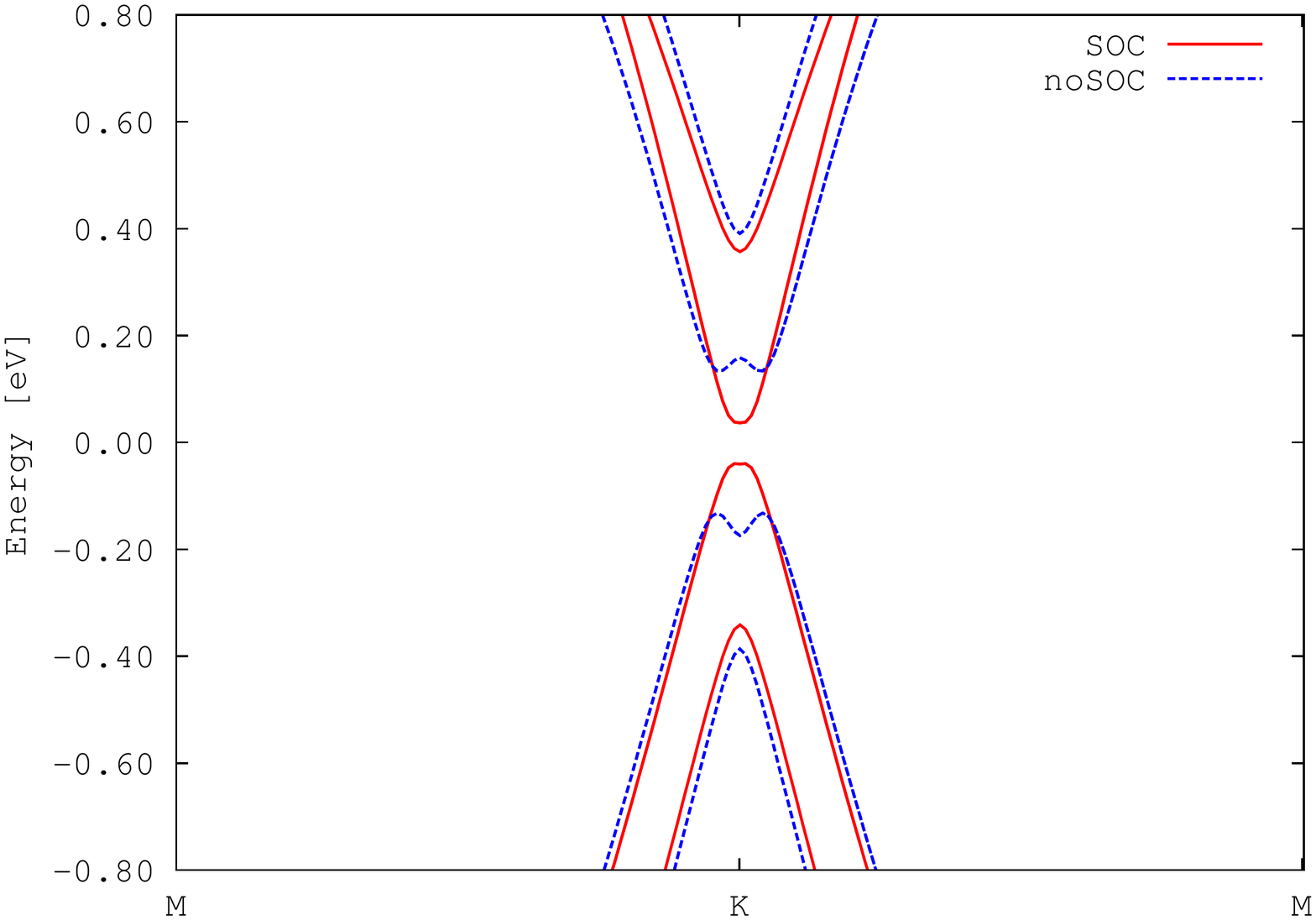}
    \end{minipage}
  \end{center}
  \caption{
    {\label{fig:elect-BS-SOC}}
    Low-energy DFT-LDA electronic band-structure of AB-stacked bilayer graphene.
    (a) Dispersions with different intensities of an applied external interlayer electric field, calculated with spin-orbit coupling.
    (b) Dispersions of a biased system ($E = 0.20$ eV/\AA) with and without spin-orbit coupling included.
    Field strengths are given in eV/\AA.    }
\end{figure}

Figure \ref{fig:elect-BS-SOC}.a shows the electronic band-structure when spin-orbit coupling (SOC) is included in the calculations.
%In the presence of strong RSOC, and for sufficiently short-range electron-electron interactions, the system minimises its energy by adopting broken-symmetry states (mostly those which break rotational symmetry) in the limit of low densities \cite{PhysRevB.85.035116}.
%These instabilities occur due to the energy dispersion having a minimum in a region of momentum-space which is bounded by two concentric circles with finite radius (annuli) \cite{PhysRevB.91.155423}.
%Moreover, distortions to the Fermi surface, resulting from a momentum-space change in the Fermi radius (a Pomeranchuk instability) can reduce lattice the lattice symmetry and lead to spontaneous longitudinal currents \cite{PhysRevB.91.155423}.

In the present study, we find that the bulk gap decreases when SOC is turned on, and then increases gradually for increasing field intensities (Figure \ref{fig:elect-BS-SOC}.a)
Under large filed intensities ($\sim$0.4 eV/\AA), the energy dispersion recovers the Mexican hat structure, since the instability occurring at the Fermi-surface competes with the SOC interaction; the energy interaction between the layers balances the coupling interaction.

Moreover, Ref. \cite{PhysRevLett.107.256801} reported that the gap vanished as the SOC parameter increases, and that on further increasing the coupling parameter it then reopens with a behaviour characteristic of a band inversion, thus suggesting a topological phase transition \cite{PhysRevLett.107.256801}.
However, since the model employed in \cite{PhysRevLett.107.256801} is different from the computations carried out for the presented work, a direct comparison between the two sets of results is not straightforward.

\subsection{Structural Instabilities of Bilayer Graphene - Lattice-Dynamics}

\begin{figure}
  \includegraphics[width=12cm]{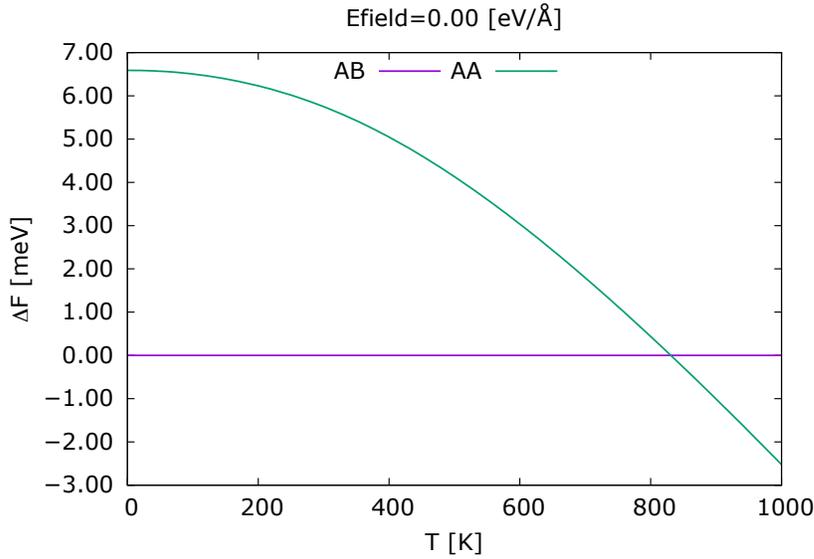}
  \caption{
    Relative free energy (Helmoltz) of the two graphene stacking environments, AB and AA, with no external electric-bias.
    The AB arrangement is calculated to be the most energetically-stable structure up to $\sim$800 K.
    }
  \label{fig:free-energy}
\end{figure}

\indent\indent
Figure \ref{fig:free-energy} shows the constant-volume (Helmoltz) free energy of the AB and AA bilayer systems calculated without an applied bias.
The energies are referenced to the lowest energy structure, which in the present calculations is the AB system.
Our calculations indicate that the AA system is energetically unstable with respect to the AB phase up to approx. 800 K, above which the AA stacking becomes lower in energy.

\begin{figure}
  \begin{center}
    \includegraphics[width=15cm]{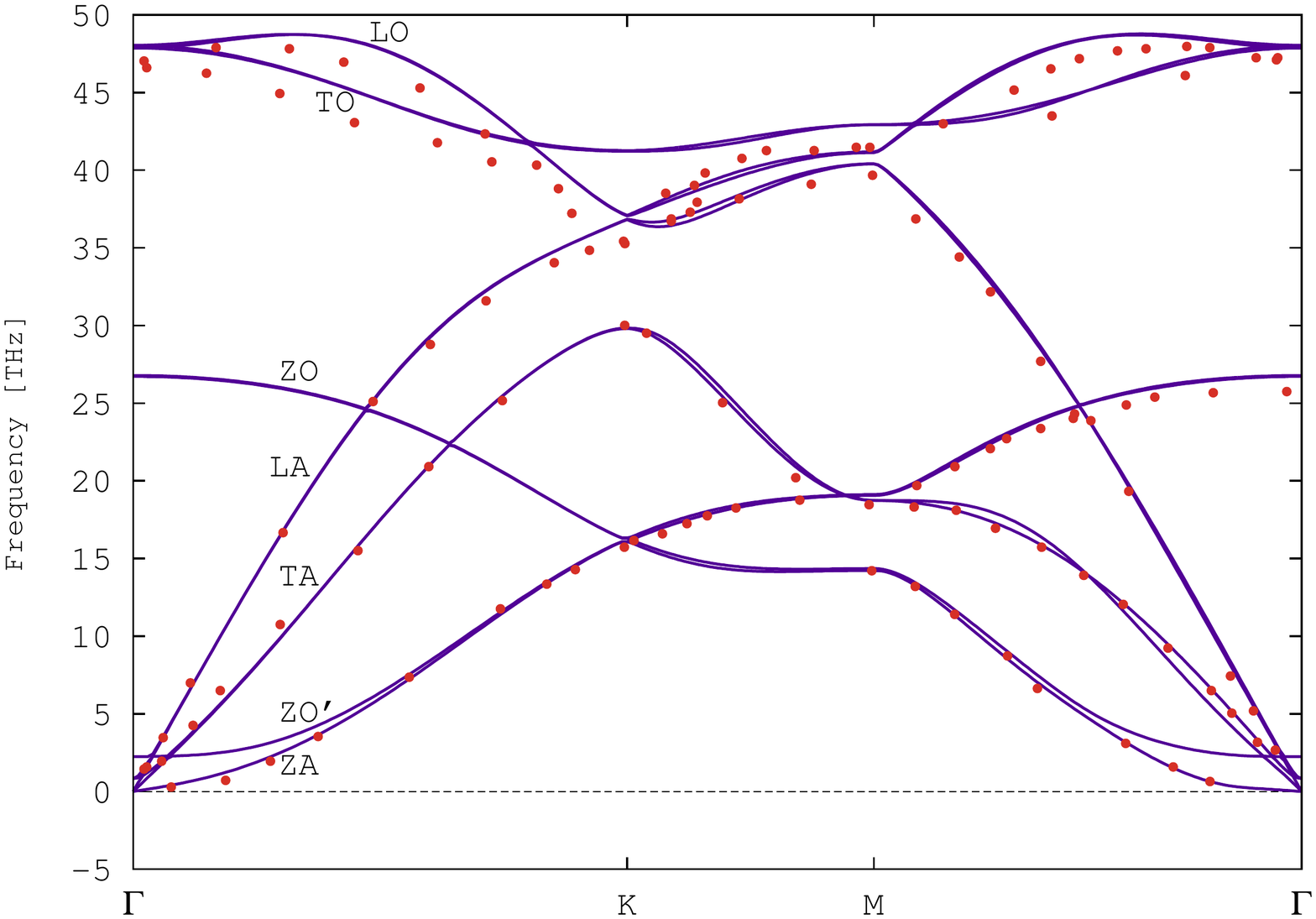}
    \caption{
      Phonon dispersion of AB-stacked bilayer graphene computed within the harmonic approximation (solid line).
      The unit cell contains four carbon atoms, leading to three acoustic (A) and $3N-3 = 9$ optical (O) phonon branches.
      The calculated dispersions are compared to the in-plane phonon dispersion of graphite obtained from inelastic x-ray scattering \cite{PhysRevB.76.035439} (red dots).
      The phonon branches are marked with the labels assigned to the $\Gamma$-point phonons.
      }
    \label{fig:bs-exp}
  \end{center}
\end{figure}

The calculated in-plane phonon dispersion agrees well with the experimental measurements on graphite presented in \cite{PhysRevB.76.035439} (Figure \ref{fig:bs-exp}), apart from a small shift of the higher-frequency TO and LO modes.
LDA calculations frequently overestimate the energies of higher-frequency phonons, but despite this difference the characteristic features of the phonon dispersion are well reproduced.

\begin{figure}
  \begin{center}
    \begin{minipage}{7.5cm}
      \begin{flushleft}
      ZA
      \end{flushleft}
      \includegraphics[width=10cm]{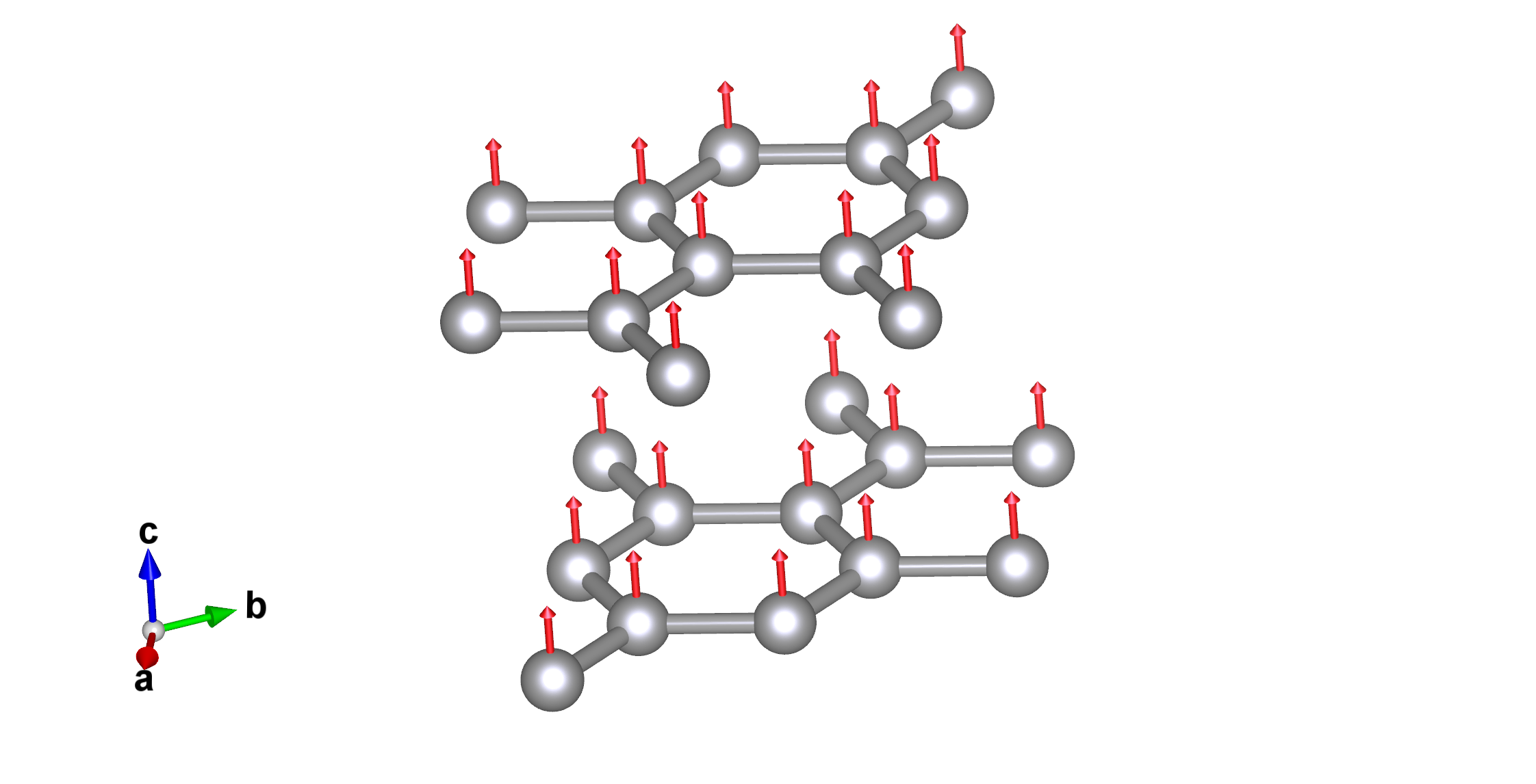}
    \end{minipage}
    \begin{minipage}{7.5cm}
      \begin{flushleft}
      ZO'
      \end{flushleft}
      \includegraphics[width=10cm]{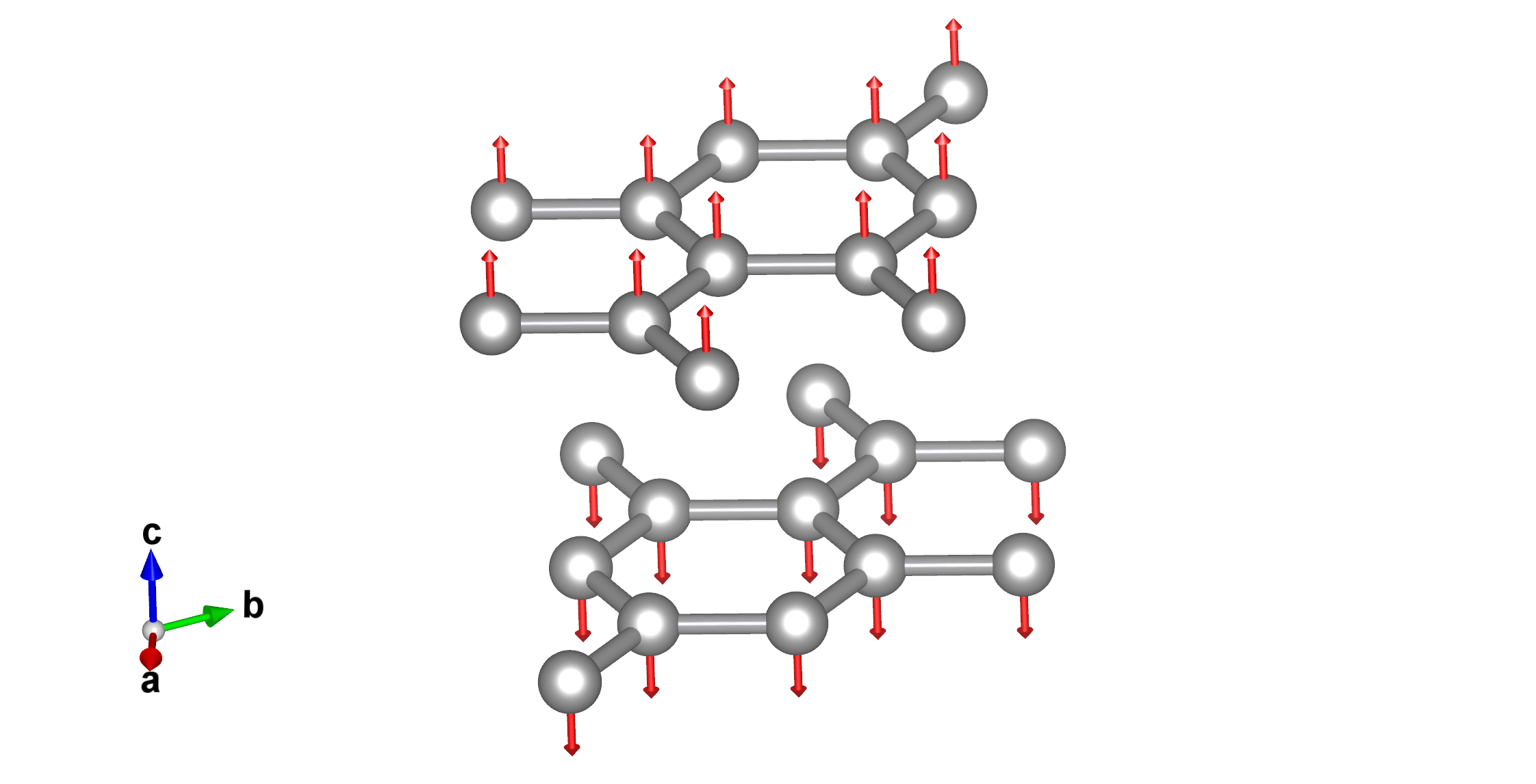}
    \end{minipage}
  \begin{center}
  TA
  \end{center}
    \begin{minipage}{7.5cm}
      \includegraphics[width=9cm]{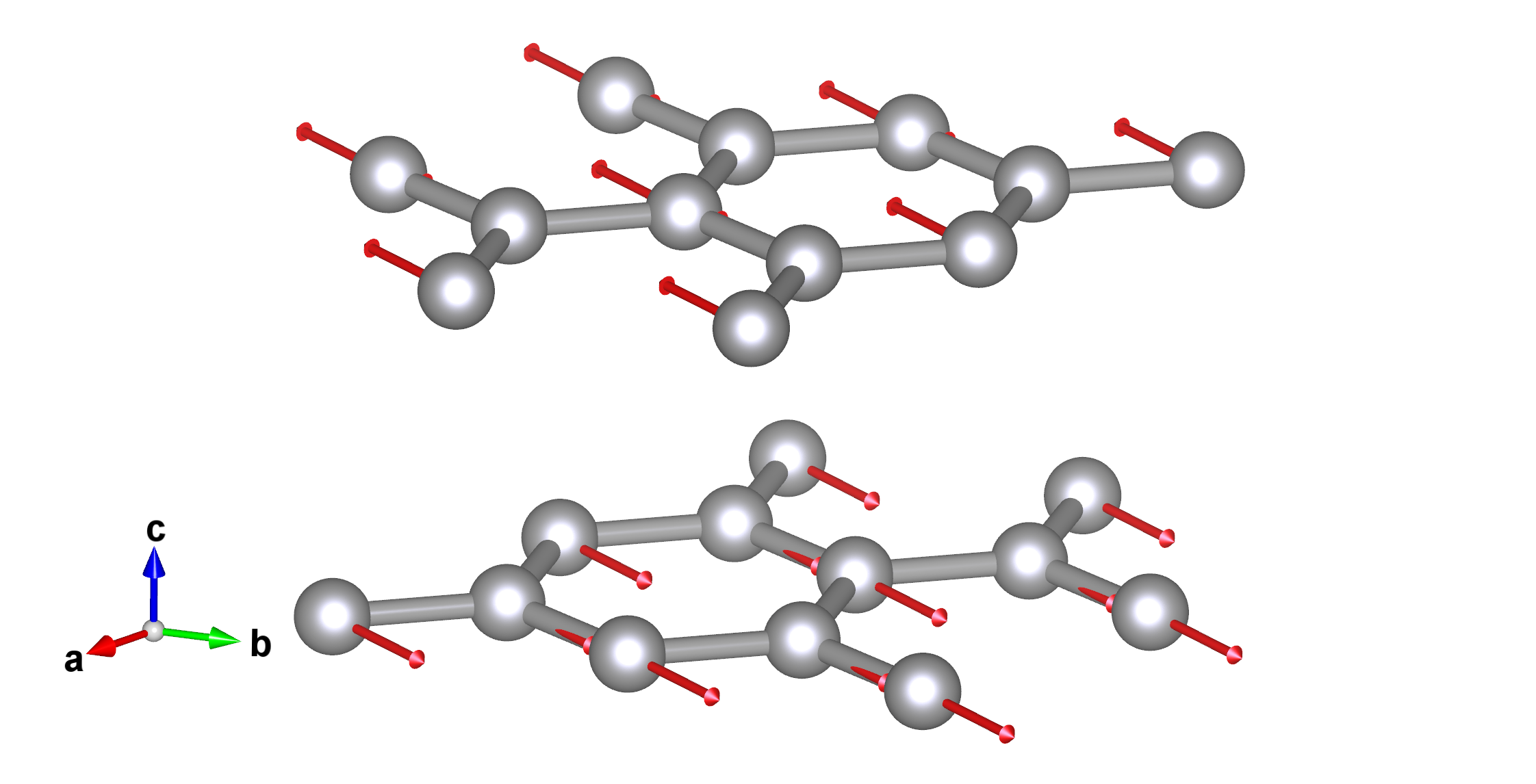}
    \end{minipage}
    \begin{minipage}{7.5cm}
      \includegraphics[width=9cm]{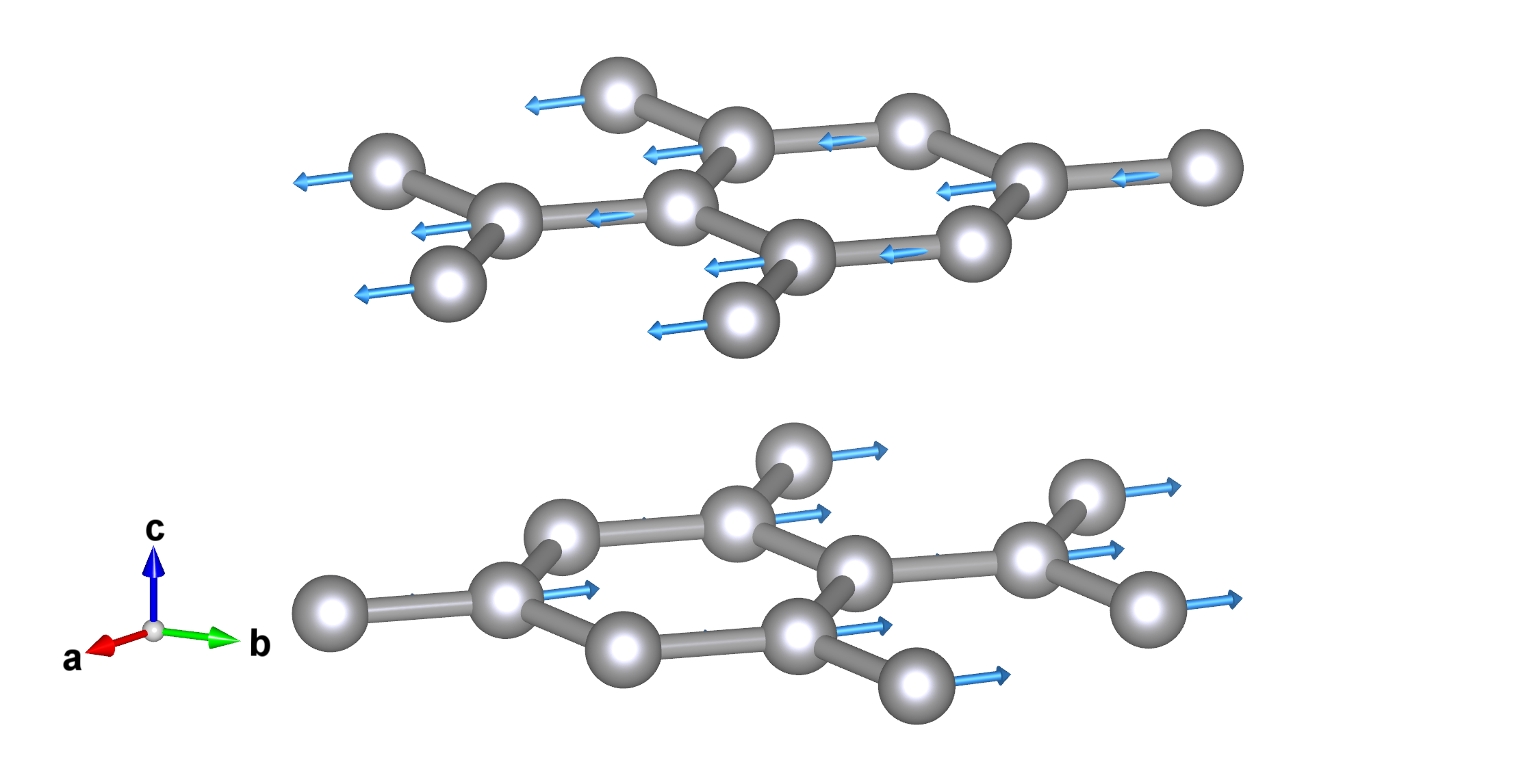}
    \end{minipage}
  \end{center}
 \caption{
  \label{fig:eigenvec}
  Eigenvectores corresponding to the vibrations in AB-stacked bilayer graphene.
  ZA and ZO' correspond to the out-of-plane vibrations, while TA denotes the degenerate in-plane transverse-acoustic modes.
  }
\end{figure}

At low \textbf{q}-vectors, the in-plane transverse acoustic (TA) and longitudinal acoustic (LA) modes show linear dispersions.
Moreover, while the doubly-degenerate LA mode has zero frequency at the $\Gamma$-point, the TA mode (also known as the shear-mode) has a non-zero frequency at the zone-center \cite{Cocemasov} ($\nu=0.82$ THz) (Figure \ref{fig:eigenvec}).

The ZA mode is the flexural acoustic mode, which corresponds to out-of-plane, in-phase atomic displacements.
In contrast to the TA and LA modes, the ZA branch shows a parabolic dispersion, i.e. $\nu \sim \mathbf{q}^2$, close to the $\Gamma$-point, indicating a low group velocity \cite{ScientRepts.7.43956.2017} and being a characteristic feature of layered materials \cite{Cocemasov, ScientRepts.7.43956.2017}.
The existence of a flexural mode is also a signature of 2D systems, and in particular is a mode which is typically found in graphene-like systems.
Since the long-wave flexural mode has the lowest frequency, it is the easiest to excite \cite{jiang-jpcom-27-2015}.

At slightly higher frequencies, the out-of-plane ZO' mode (Figure \ref{fig:eigenvec}) can be observed, which corresponds to interlayer motion along the $Z$-axis (a layer-breathing mode).
The other out-of-plane optic modes are characterised by the doubly degenerate ZO branch.
At the $\Gamma$ point, the interlayer coupling causes the LO and TO modes to split into two doubly-degenerate branches, both of which correspond to in-plane relative motion of atoms.
With the exception of the ZA and ZO' modes, all the frequency branches have symmetry-imposed degeneracy at $\Gamma$ (Figure \ref{fig:bs-exp}).

% JMS: It would be technically more correct to use $i$ instead of negative values for imaginary frequencies.
% JMS: The behaviour of the ZA mode in Fig. 8 is probably an interpolation artefact.

\begin{figure}
  \begin{center}
    \begin{minipage}{9.5cm}
      \includegraphics[width=12cm]{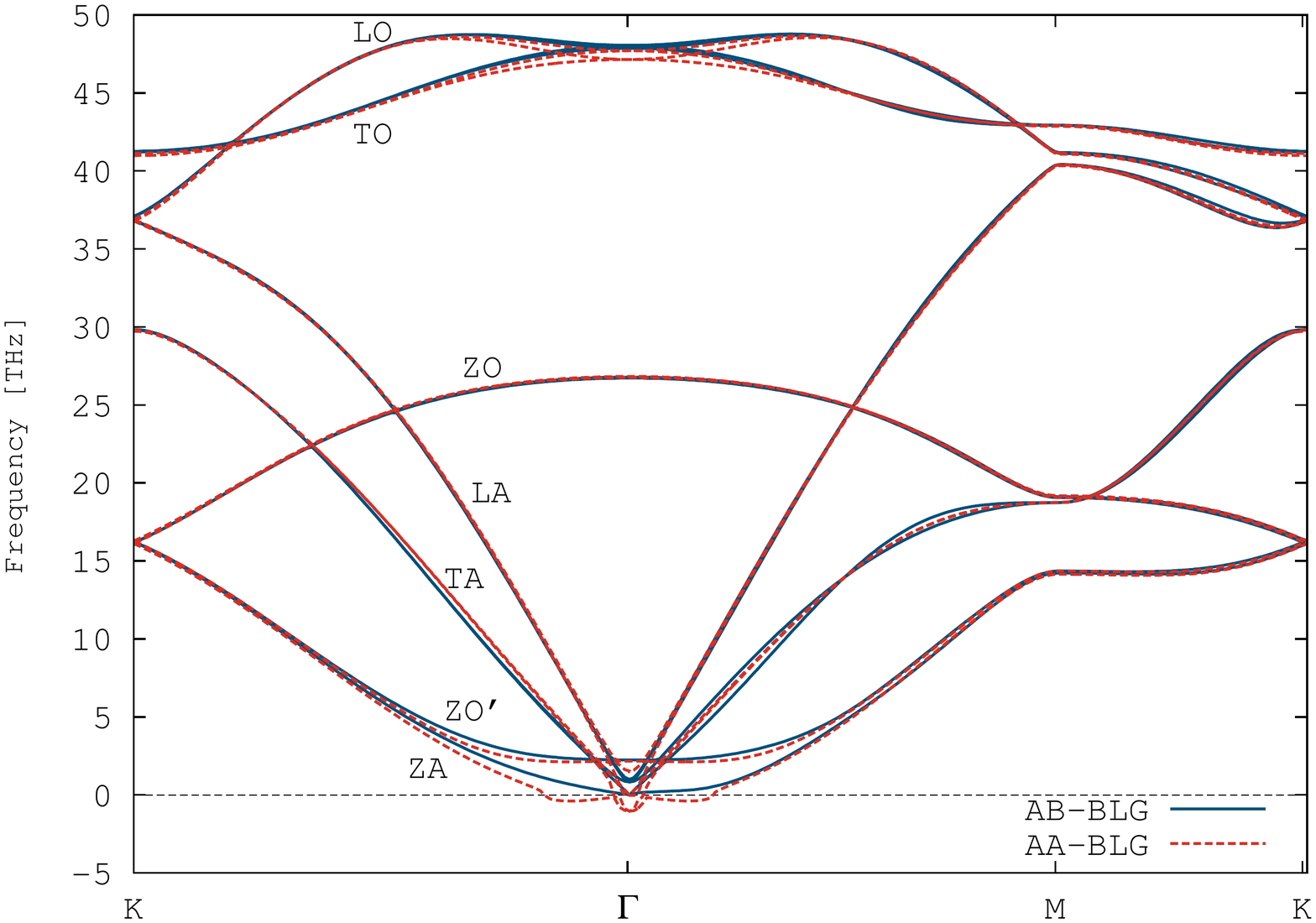}
    \end{minipage}
    \begin{minipage}{3cm}
      \includegraphics[width=12cm]{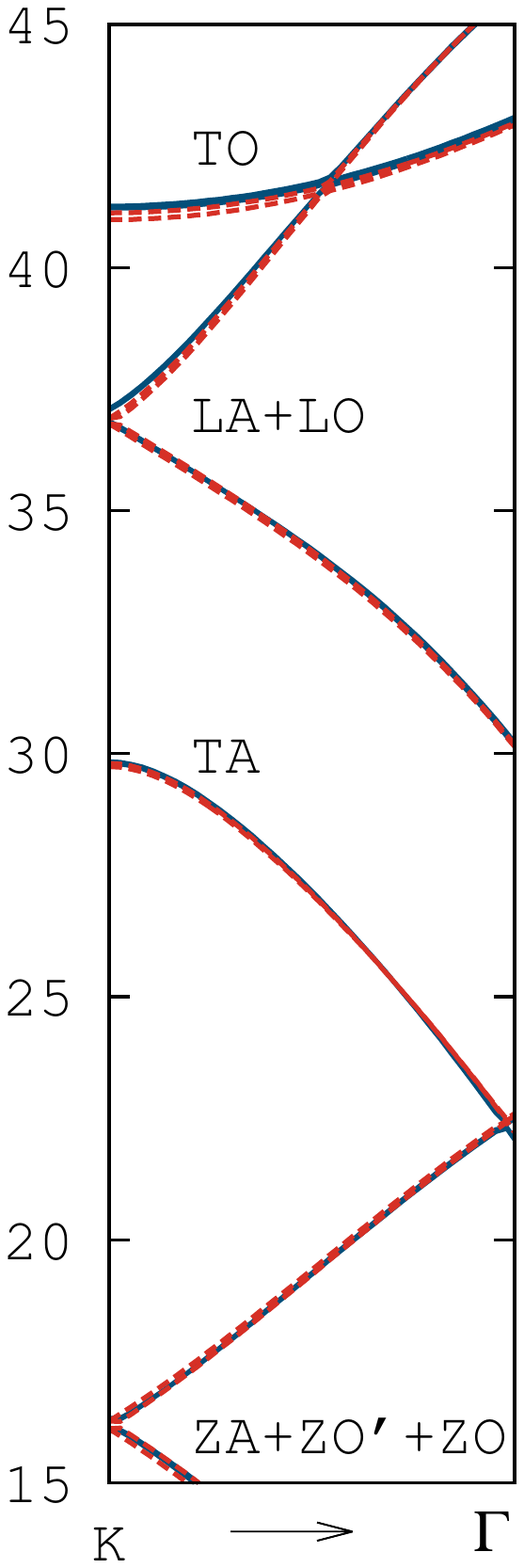}
    \end{minipage}
  \end{center}
  \caption{
    \label{fig:ABAA-KG}
    Phonon dispersions of the AB- and AA-stacked bilayer graphene systems computed with the harmonic approximation (blue solid and red dashed lines, respectively).
    The right-hand panel shows the dispersion along the K-$\Gamma$ path.
    The phonon branches are denoted by the symbols of the $\Gamma$-point phonons, several of which become degenerate at the K-point.
    }
  \label{fig:bsAAAB}
\end{figure}

Figure \ref{fig:ABAA-KG} compares the phonon dispersions of the two stacking modes.
Both stacking configurations have similar mode characters, although differences emerge at the zone-centre.

For the AA system, a small phonon instability is observed at the $\Gamma$-point, which is denoted by an imaginary mode ($\nu =i 1.04$ THz).
This indicates that the AA-system is dynamically unstable, and prefers to adopt the AB-stacking configuration, in accordance with the free energies in Figure \ref{fig:free-energy}.
As expected, the imaginary mode is a TA branch, which corresponds to the shear displacement of the layers with respect to one another.
The ZA mode also shows instabilities in the vicinity of the zone-center, but has zero frequency at $\Gamma$.

The ZO' breathing mode of AA-stacked bilayer graphene is located in a similar frequency range to the corresponding mode in AB graphene, at $\nu=2.16$ and $\nu=2.25$ THz, respectively.
The biggest frequency differences are observed for the TO modes, which in the AB system occurs at higher frequency than in the AA configuration, with 0.72 THz of difference.
This is partly because the LO/TO is larger in the AA than the AB system (0.57 and 0.18 THz, respectively).

Table \ref{table:modes} presents a summary of the zone-centre frequencies for the two stacking configurations. %, and the results are in accord with literature values.

% JMS: This paragraph is left hanging - did you mean to say that your LDA calculations are probably more accurate (or words to that effect)?

We note that the AA phonon dispersion does not correspond to that in \cite{PhysRevB.88.035428}, where, in contrast to the present results, imaginary frequencies are not observed (with calculations carried out using the Born-von-Karman model of lattice dynamics for in-plane atomic coupling and the Lennard-Jones potential for interlayer coupling \cite{PhysRevB.88.035428}).

\begin{table}
  \caption{
    \label{table:modes}
    Frequencies (THz) of the $\Gamma$-point phonon modes in AB- and AA-stacked bilayer graphene.
    }
  \begin{indented}
    \item[]
    \begin{tabular}{@{}llllllll}
      \br
      Mode &    ZA & ZO' & TA & LA & ZO & TO & LO \\
      AB &   0.00 & 2.25 & 0.82 & 0.00 & 26.72 & 47.86 & 48.04\\
      AA & 0.00 & 2.16 & $\imath$ 1.04 & 0.00 & 26.82 & 47.14 & 47.71 \\
      \br
    \end{tabular}
  \end{indented}
\end{table}

The branches which originate from the out-of-plane modes at the $\Gamma$-point, i.e. ZA, ZO' and ZO, become degenerate at the K-point (Figure \ref{fig:ABAA-KG})).
The in-plane LO and LA phonon branches also meet at the $K$-point, giving rise to a doubly-degenerate phonon band.
It is also noteworthy that the dispersions of the out-of-plane modes behave linearly around the $K$-point in AA-BLG, whereas those in AB-BLG show a parabolic-like dispersion similar to that suggested in \cite{PhysRevB.88.035428}.
Features in the electronic spectra near the K-point in the two BLG systems are therefore also reflected in the phonon spectra (c.f. Figures \ref{fig:elect-BS} and \ref{fig:ABAA-KG}).

\begin{figure}
  \begin{center}
    \begin{minipage}{12cm}
      \begin{flushleft}
      a)
      \end{flushleft}
      \includegraphics[width=12cm]{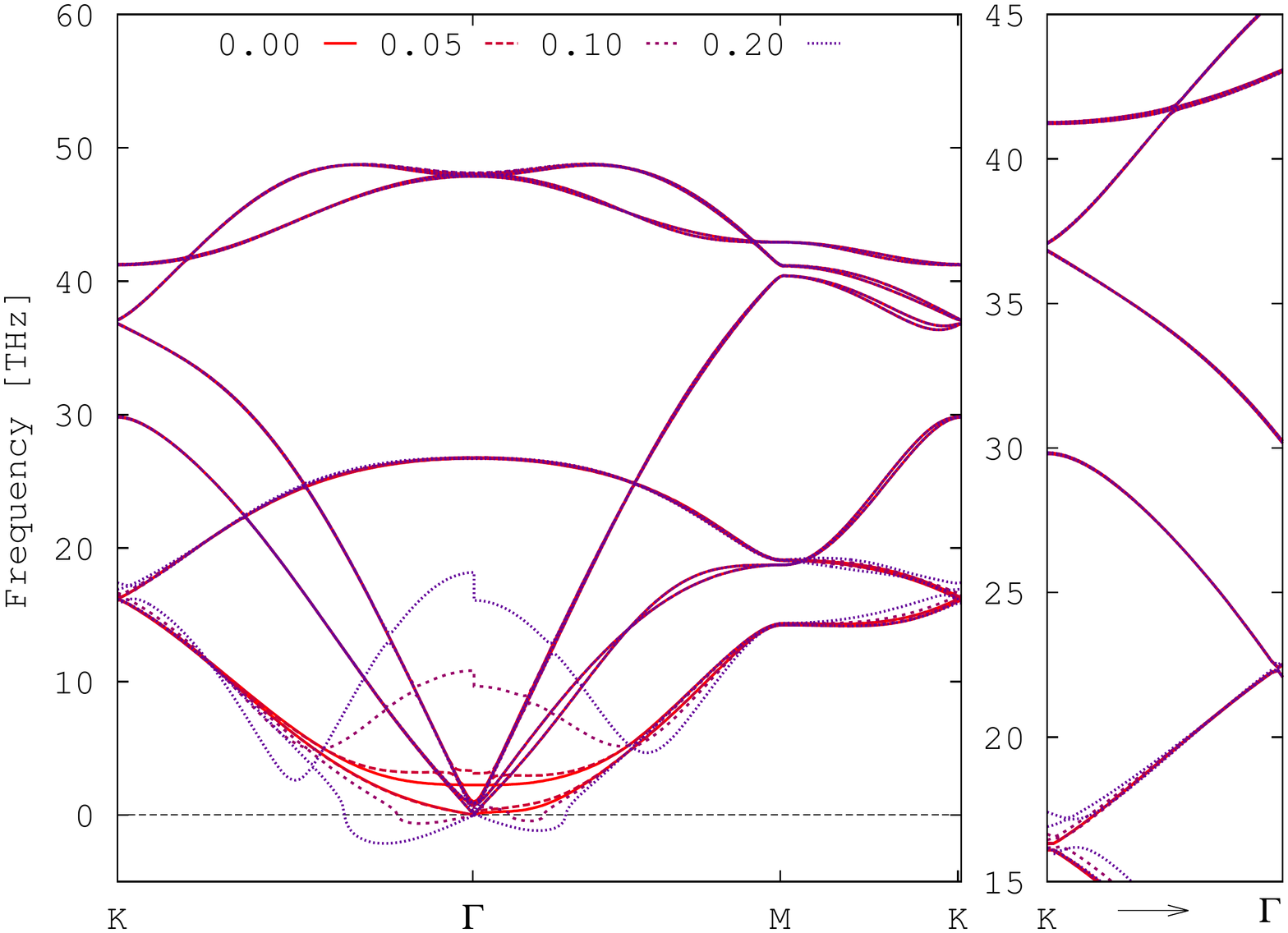}
    \end{minipage}
    \begin{minipage}{12cm}
      \begin{flushleft}
      b)
      \end{flushleft}
      \includegraphics[width=10cm]{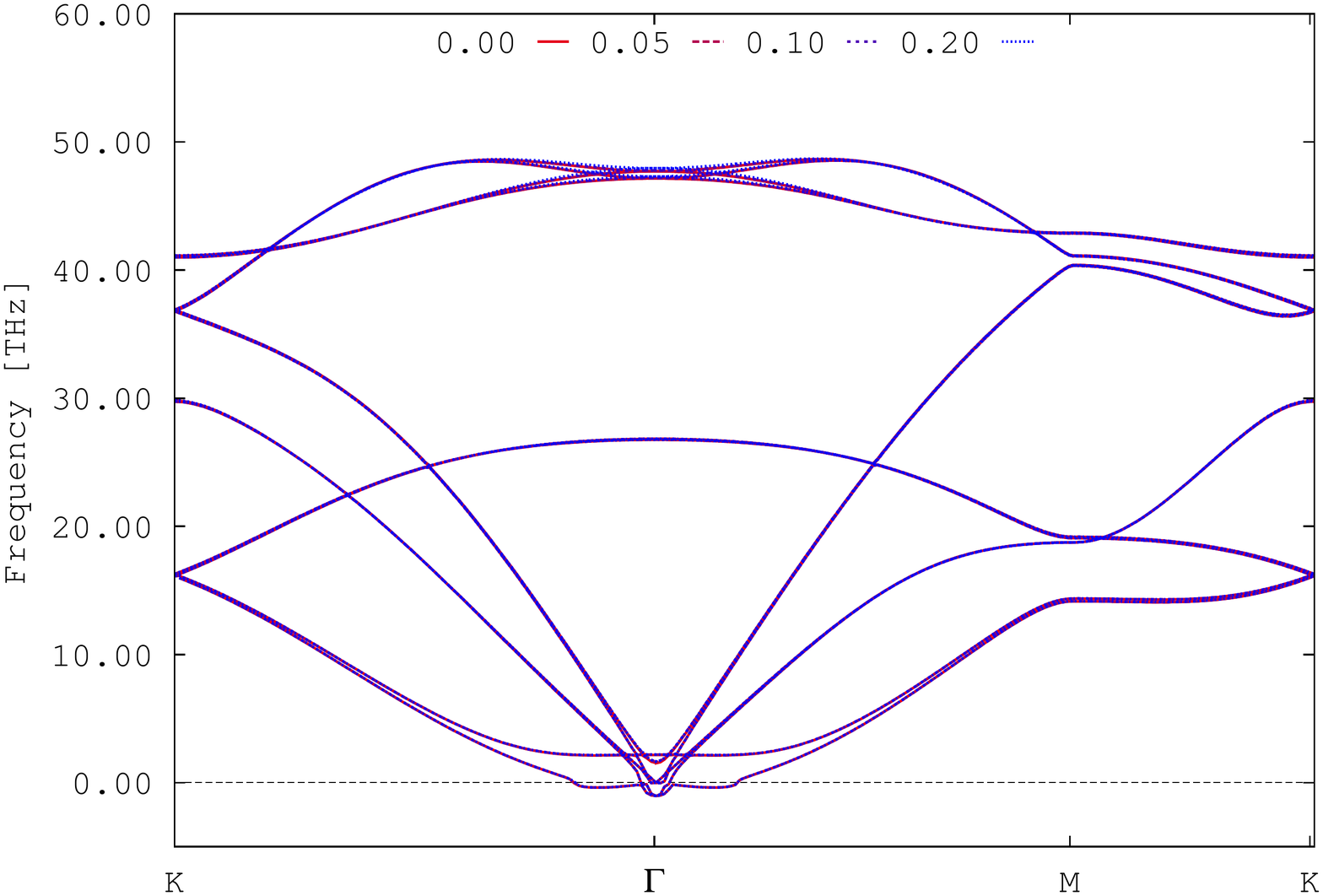}
    \end{minipage}
  \end{center}
  \caption{
    {\label{fig:phonon-BS}}
    Harmonic phonon dispersions of the AB- (a) and AA-stacking (b) configurations of bilayer graphene under different applied bias.
    The right-hand panel in (a) shows the dispersion along the K-$\Gamma$ segment.
    The AA system shows doubly-degenerate imaginary modes at the zone-center, which is consistent with the alternative AB stacking being the most favourable arrangement.
    Non-analytical corrections have been applied to the dispersions of both systems.
    The ZA and ZO' modes on the AB-BLG change significantly under bias, whereas the applied field has comparatively little effect on the dispersion of the AA-BLG system.
    Electric fields are given in eV/\AA.
    }
\end{figure}

Further lattice-dynamics calculations were carried out to investigate the effect of electric fields on the phonon dispersions (Figure \ref{fig:phonon-BS}).
Non-analytical corrections to the dynamical matrix at $q \rightarrow 0$ were considered in all calculations.
We find that the dispersion of the AA system is relatively insensitive to the applied external bias, and that for all applied fields the $\Gamma$-point instability persists.

% JMS: Out of interest, what are the Born charges and dielectric constants of the two models, and how are they affected by the applied bias (if you tested this)?
% JMS: Also, I am not 100 % convinced that those imaginary modes in the AB dispersion are not due interpolation artefacts.

In comparison, the low-frequency branches of the AB band-structure show a significant response to the field (Figure \ref{fig:phonon-BS}.a).
This effect results from the inclusion of non-analytical corrections; when these corrections are not included, the dispersions are relatively unaffected by the bias.
The layer-breathing mode (ZO') displays a discontinuity at the $\Gamma$-point, with different frequencies for different directions of approach.
Moreover, the flexural-acoustic (ZA) mode shows instabilities in the vicinity of the zone-centre, but continue to show zero frequency at the $\Gamma$-point.
Since the long-wave flexural mode has the lowest frequency, it is the easiest to excite \cite{jiang-jpcom-27-2015} and is therefore more sensitive to the bias.
At the K-point (Figure \ref{fig:phonon-BS}, inset), as occurs for the electronic band-structure the degeneracy of the out-of-plane modes split, with the magnitude of the splitting depending on the size of the applied bias.

\section{Conclusions}

\indent\indent
In summary, we have performed a detailed first-principles study of the effect of applied fields on the electronic structure and lattice dynamics of bilayer graphene.

Application of an external field to the AB-stacked bilayer graphene system leads to drastic changes in the electronic properties, leading to the opening of the gap and asymmetry in the dispersion.
This in turn induces in-plane inhomogeneities in the charge distribution on the sublattices, and the Coulomb interaction between electrons will thus cause a potential difference between the layers.
Our results therefore show that the electron density can be controlled by tuning the band-gap width and dispersion asymmetry.

Spin-orbit coupling has a significant effect on the dispersion as short-range electron-electron correlations become important.
The Mexican-hat structure disappears under low bias, and the energy gap decreases.
At larger field strength, the asymmetry in the dispersion persists, since the energy scale set by the Fermi-surface instability is minimized.

On the other hand, the electronic structure of the AA system is relatively stable under bias.

As for its electronic structure, applied fields cause the phonon dispersions of the AB-stacked system to change significantly when non-analytical corrections for long-range Coulomb interactions are taken into account.
These corrections mainly affect the lower-frequency out-of-plane  ZA and ZO' modes.
The phonon dispersion of the AA system shows degenerate imaginary modes at the $\Gamma$ point, indicating the presence of a phonon instability.
The dispersion of this stacking configuration is relatively insensitive to bias and does not change significantly in response to an applied field.

In order to obtain better consistency with available literature, we would need to go beyond LDA functional.
The ground-state is likely to have additional broken-symmetry configurations, and the lifting of spin and valley degeneracies may depend on long-range fluctuations, effects which are not well captured by local DFT functionals.
For example, in the literature it has been observed that the AA stacking configuration may be stabilised by an excitonic gap \cite{PhysRevLett.109.206801}.
To study such effects, one would need to resort to the two-body Green function method (Bethe-Salpeter equation), a possibility which we are currently exploring.

\ack

\indent\indent
This work was supported by EPSRC Programme Grants (nos. EP/K004956/1 and EP/K016288/1) and the ERC (Grant No. 277757).
We acknowledge use of the ARCHER supercomputer through the PRACE Research Infrastructure (award no. 13DECI0313).
The authors also acknowledge computing support from the University of Bath Computing Services, which maintains the Balena HPC cluster.

\clearpage
\appendix

\section{Tight-Binding Hamiltonian}

\label{appendix-sec1}

\indent\indent
To study the effects of an applied electric field on bilayer graphene, we begin by describing pure bilayer graphene system with Bernal stacking, considering only the $t_z$ interlayer hopping amplitude, restricted to the nearest-neighbour carbon atoms.
An electric bias $V=eEd$ is applied in the direction perpendicular to the layers, where $e$ is the electron charge, $E$ the applied electric field, and $d$ the interlayer spacing \cite{pogorelov-prb-92-2015}.
The Fermi operators, written as:

\begin{eqnarray}
 \label{eq:fermi_op}
 a_{\mathbf{k}j}=N^{-1} \sum_n e^{\imath \mathbf{k} \cdot \mathbf{n}}  a_{\mathbf{n}j} \qquad
 b_{\mathbf{k}j}=N^{-1} \sum_n e^{\imath \mathbf{k} \cdot \mathbf{n}}  b_{\mathbf{n}j}
\end{eqnarray}

\noindent\noindent
represent plane-wave states with momentum $\mathbf{k}$.
The indices $j$ refer to the layers in the $n$th unit-cell, $a_{\mathbf{n}j}$ and $b_{\mathbf{n}j}$ are operators referring to A- and B-type sites in the lattice (figure \ref{fig:BLG}), and $N$ is the number of cells in a layer.
A $4-$spinor is formed by:

\begin{equation}
  \label{eq:spinors}
  \psi_{\mathbf{k}}^\dagger= \left(a_{\mathbf{k}1}^\dagger, b_{\mathbf{k}1}^\dagger, a_{\mathbf{k}2}^\dagger, b_{\mathbf{k}2}^\dagger \right).
\end{equation}

\noindent\noindent
The Hamiltonian is given in the second-quantized form by:

\begin{equation}
  \label{eq:un_hamil}
  H_0 = \sum_\mathbf{k} \psi_{\mathbf{k}}^\dagger \hat{H}_\mathbf{k} \psi_{\mathbf{k}}
\end{equation}

\noindent\noindent
where the matrix has the form:

\begin{equation}
  \hat{H}_\mathbf{k} =
    \left(
      \begin{array}{cccc}
                      \frac{V}{2} & \gamma_\mathbf{k} &                         0 &               t_z \\
        \gamma_\mathbf{k}^\dagger &       \frac{V}{2} &                         0 &                 0 \\
                                0 &                 0 &             - \frac{V}{2} & \gamma_\mathbf{k} \\
                              t_z &                 0 & \gamma_\mathbf{k}^\dagger & - \frac{V}{2}
      \end{array}
    \right)
  \label{eq:Hbilayer}
\end{equation}

\noindent
The term $\gamma_\mathbf{k} = t \sum_{\boldsymbol \delta} e^{\imath \mathbf{k} \cdot \boldsymbol \delta}$ is related to the in-plane hopping amplitude, $t$, over the nearest-neighbour vectors $\boldsymbol \delta$ and defined as $t \approx 3$ according to experimental and first principles calculations suggested in Refs. \cite{castro-graphite} and \cite{charlier_prb_1991}). The variable $\xi_\mathbf{k}$ is defined as $ \hbar v_F(\textbf{k}-K)$, with the Fermi velocity defined as $v_F=3ta/2\hbar\approx 10^{-6}$ms$^{-1}$, and $a$ being the intra-layer distance between next-neighbour atoms of different sublattices and $\hbar$ the Planck constant \cite{pogorelov-prb-92-2015}. The interlayer hopping amplitude, $t_z$, is related to $t$ by the relation $t_z = t/10$ \cite{castro-graphite}.

In the vicinity of the Dirac points $\pm K$ is defined by $\pm (4 \pi/3\sqrt{3}a,0)$ and $\gamma_\mathbf{k} \approx \xi_\mathbf{k} e^{\imath \varphi_\mathbf{k} }$, with $\varphi_\mathbf{k}= \tan^{-1}{k_y/(k_x-K_x)}$ \cite{pogorelov-prb-92-2015}.
After diagonalizing the matrix in Eq. \ref{eq:Hbilayer}, and considering that the dispersion in the vicinity of the Dirac point is defined along a circle around each Dirac point, $\xi_\mathbf{k} \equiv \xi$, with maximum radius defined as $\hbar v_F \sqrt{K/a}$ \cite{pogorelov-prb-92-2015}, we obtain for the four bands:

\begin{eqnarray}
  \label{eq:eigen_energies}
  \varepsilon_{1}(\xi) &=& \pm \frac{1}{2} \sqrt{2t_{z}^2+V^2 + 4 \xi^2+2\sqrt{t_{z}^4 + 4(t_{z}^2 + V^2)\xi^2}} \nonumber \\
  \varepsilon_{2}(\xi) &=& \pm \frac{1}{2} \sqrt{2t_{z}^2+V^2 + 4 \xi^2-2\sqrt{t_{z}^4 + 4(t_{z}^2 + V^2)\xi^2}}
\end{eqnarray}

\noindent\noindent
where $\varepsilon_{1}$ and $\varepsilon_{2}$ refer to the external and internal bands, respectively \cite{pogorelov-prb-92-2015}. %The radial variable, $\xi$ is evaluated near the Dirac point with the choice of $V=2t_z$ as defined in Ref. \cite{castro-graphite, pogorelov-prb-92-2015}.

The bias-controlled gap will therefore result in:

\begin{eqnarray}
  \label{eq:gap}
  \varepsilon_\textrm{g}= \pm \frac{V}{ 2 \sqrt{1+(\frac{V}{t_z})^2} }.
\end{eqnarray}

For the AA-BLG, the calculations are similar, although the matrix takes the form:

\begin{equation}
  \label{eq:Hbilayer-AA}
  \hat{H}_\mathbf{k} =
  \left(
    \begin{array}{cccc}
                    \frac{V}{2} & \gamma_\mathbf{k}  &                       t_z &                 0 \\
      \gamma_\mathbf{k}^\dagger &        \frac{V}{2} &                         0 &               t_z \\
                            t_z &                  0 &              -\frac{V}{2} & \gamma_\mathbf{k} \\
                              0 &                t_z & \gamma_\mathbf{k}^\dagger & -\frac{V}{2}
    \end{array}
  \right)
\end{equation}
\noindent\noindent and hence the eigenenergies will result in

\begin{eqnarray}
  \label{eq:eigen_energies}
  \varepsilon_{1}(\xi) &=& \pm \frac{1}{2}\sqrt{4 t_z^2 + V^2} + \xi \nonumber \\
  \varepsilon_{2}(\xi) &=& \pm \frac{1}{2}\sqrt{4 t_z^2 + V^2} - \xi.
\end{eqnarray}

\begin{figure}[!]
  \begin{center}
    \begin{minipage}{7.0cm}
      \begin{flushleft}
      a)
      \end{flushleft}
      \includegraphics[width=12cm]{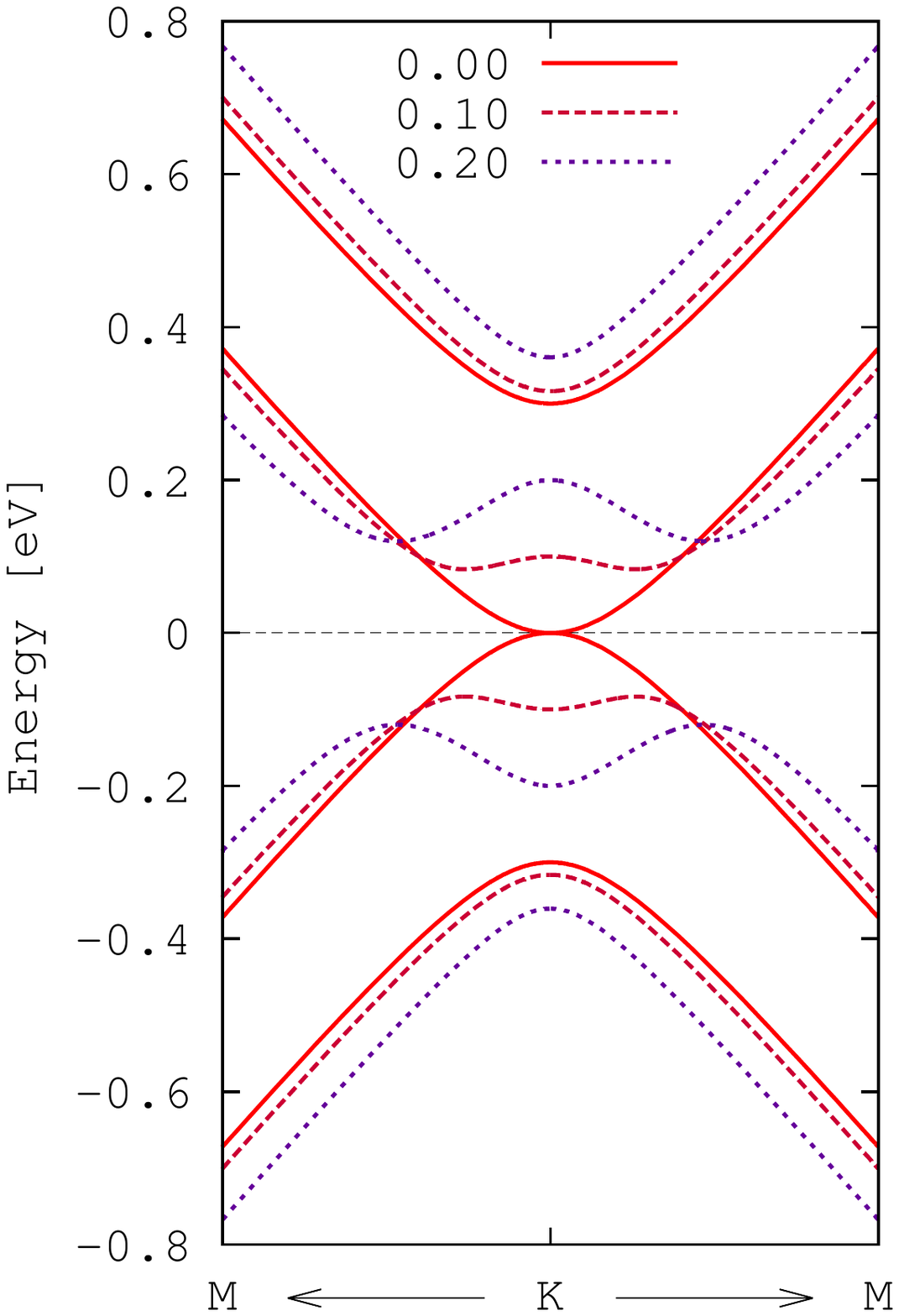}
    \end{minipage}
    \begin{minipage}{7.0cm}
      \begin{flushleft}
      b)
      \end{flushleft}
      \includegraphics[width=12cm]{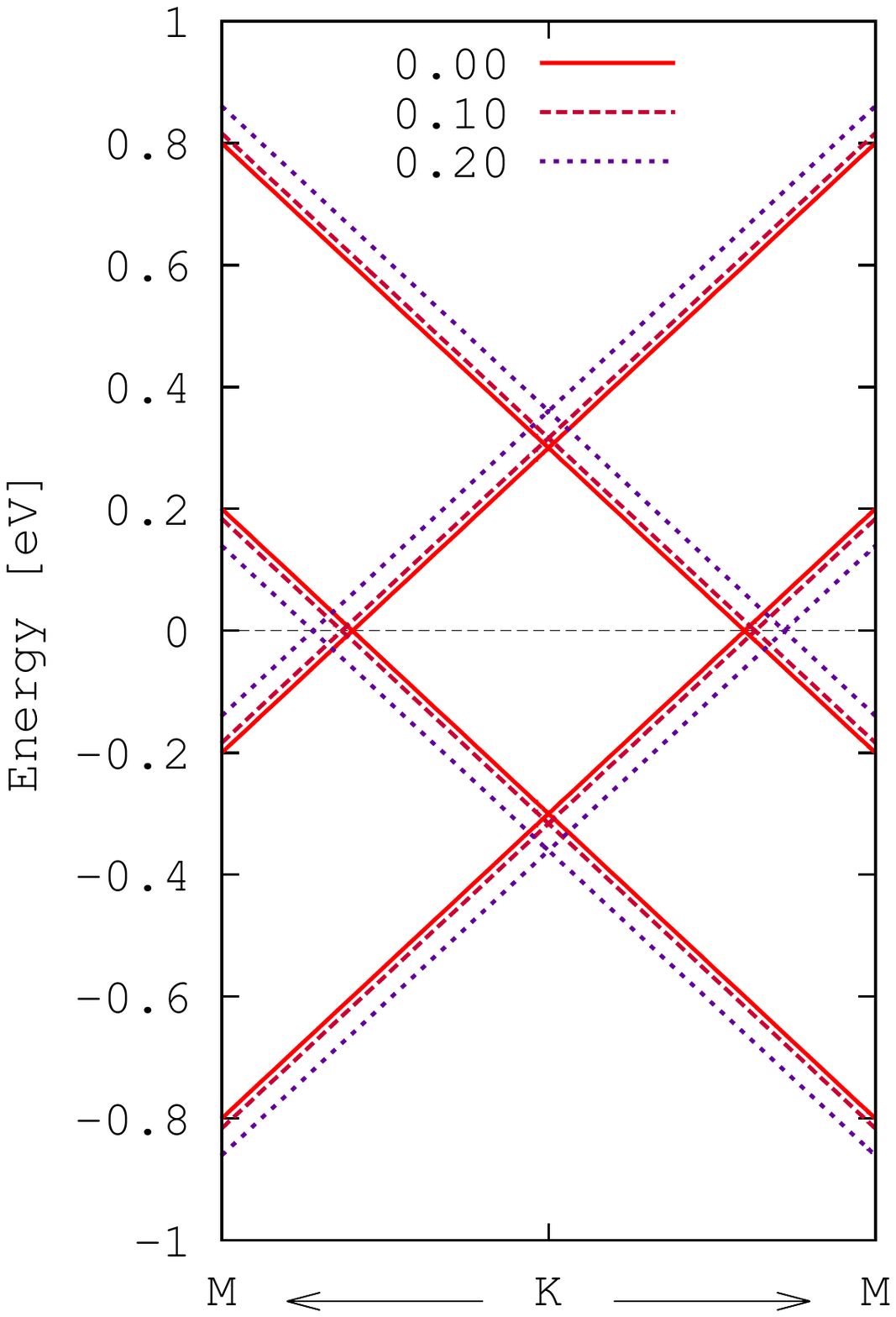}
    \end{minipage}
   \end{center}
   \caption{
    {\label{fig:elect-TB}}
    Low-energy tight-binding electronic band-structure of bilayer graphene in the AB (a) and AA (b) stacking arrangements.
    Dispersions are calculated for different intensities of $V$.
    }
 \end{figure}

\break

%\bibliographystyle{unsrt}
%\bibliography{biblo}

\begin{thebibliography}{10}

\bibitem{berger04_short}
C.~Berger, Z.~Song, T.~Li, X.~Li, A.~Y. Ogbazghi, R.~Feng, Z.~Dai, A.~N.
  Marchenkov, E.~H. Conrad, P.~N. First, and W.~A. de~Heer.
\newblock {U}ltrathin {E}pitaxial {G}raphite: 2{D} {E}lectron {G}as
  {P}roperties and a {R}oute {T}oward {G}raphene-{B}ased {N}anoelectronics.
\newblock {\em J. Phys. Chem.}, 108:19912, 2004.

\bibitem{pogorelov-prb-92-2015}
M.~C.~Santos Y.~G.~Pogorelov and V.~M. Loktev.
\newblock {E}lectric {B}ias {C}ontrol of {I}mpurity {E}ffects in {B}ilayer
  {G}raphene.
\newblock {\em Phys. Rev. B}, 92:075401, 2015.

\bibitem{Novoselov2006_bilayer_short}
K.~S. Novoselov, E.~McCann, S.~V. Morozov, V.~I. Fal'ko, M.~I. Katsnelson,
  U.~Zeitler, D.~Jiang, F.~Schedin, and A.~K. Geim.
\newblock {U}nconventional {Q}uantum {H}all {E}ffect and {B}erry's {P}hase of
  2$\pi$ in {B}ilayer {G}raphene.
\newblock {\em Nature Physics}, 2:177, 2006.

\bibitem{ginaluca}
G.~Fiori and journal = {IEEE Electron Device Lett.} volume = {30} pages =
  {1096} year =~{2009} G.~Iannaccone, title =~{{U}ltralow-{V}oltage {B}ilayer
  {G}raphene {T}unnel {FET}}.

\bibitem{kuhne}
journal = {Nature Nanot.} volume = {12} pages = {895} year =~{2017}
  M.~K\"{u}hne, title =~{{U}ltrafast {L}ithium {D}iffusion in {B}ilayer
  {G}raphene}.

\bibitem{zhao.natcomm.5.3410}
J.-Q. Huang G.-L. Tian J.-Q. Nie H.-J.~Peng M.-Q.~Zhao, Q.~Zhang and journal =
  {Nature Comms.} volume = {5} pages = {3410} year =~{2014} F.~Wei, title
  =~{{U}nstacked {D}ouble-{L}ayer {T}emplated {G}raphene for {H}igh-{R}ate
  {L}ithium–{S}ulphur {B}atteries}.

\bibitem{yan_nanophotonics_2015}
H.~Yan.
\newblock {B}ilayer {G}raphene: {P}hysics and {A}pplication {O}utlook in
  {P}hotonics.
\newblock {\em Nanophotonics}, 4:115, 2015.

\bibitem{akbari_jnano_2014}
E.~Akbari, R.~Yusof, M.~T. Ahmadi, A.~Enzevaee, M.~J. Kiani, H.~Karimi, and
  M.~Rahmani.
\newblock {B}ilayer {G}raphene {A}pplication on {NO}$_2$ {S}ensor {M}odelling.
\newblock {\em J. Nanomat.}, 2014:1, 2014.

\bibitem{Novoselov2005.Nature}
S.~V. Morozov D. Jiang M. I. Katsnelson I. V. Grigorieva S. V. Dubonos A.
  A.~Firsov K.~S.~Novoselov, A. K.~Geim.
\newblock {T}wo-{D}imensional {G}as of {M}assless {D}irac {F}ermions in
  {G}raphene.
\newblock {\em Nature}, 438:197, 2005.

\bibitem{PhysRep.648.2016}
A.~L. Rakhmanov F.~Nori A.~V.~Rozhkov, A. O.~Sboychakov.
\newblock {E}lectronic {P}roperties of {G}raphene-{B}ased {B}ilayer {S}ystems.
\newblock {\em Phys. Rep.}, 648.

\bibitem{PhysRevB.95.075438}
P.~G. Silvestrov and P.~Recher.
\newblock {W}igner {C}rystal {P}hases in {B}ilayer {G}raphene.
\newblock {\em Phys. Rev. B}, 95:075438, Feb 2017.

\bibitem{Jaeger1998}
Gregg Jaeger.
\newblock The ehrenfest classification of phase transitions: Introduction and
  evolution.
\newblock {\em Archive for History of Exact Sciences}, 53(1):51--81, May 1998.

\bibitem{0034-4885-76-5-056503}
E.~McCann and M.~Koshino.
\newblock {T}he {E}lectronic {P}roperties of {B}ilayer {G}raphene.
\newblock {\em Reports on Progress in Physics}, 76(5):056503.

\bibitem{PhysRevLett.107.256801}
Zhenhua Qiao, Wang-Kong Tse, Hua Jiang, Yugui Yao, and Qian Niu.
\newblock {T}wo-{D}imensional {T}opological {I}nsulator {S}tate and
  {T}opological {P}hase {T}ransition in {B}ilayer {G}raphene.
\newblock {\em Phys. Rev. Lett.}, 107:256801, Dec 2011.

\bibitem{PhysRevB.85.035116}
Erez Berg, Mark~S. Rudner, and Steven~A. Kivelson.
\newblock {E}lectronic {L}iquid {C}rystalline {P}hases in a {S}pin-{O}rbit
  {C}oupled {T}wo-{D}imensional {E}lectron {G}as.
\newblock {\em Phys. Rev. B}, 85:035116, Jan 2012.

\bibitem{PhysRevB.91.155423}
Jeil Jung, Marco Polini, and A.~H. MacDonald.
\newblock {P}ersistent {C}urrent {S}tates in {B}ilayer {G}raphene.
\newblock {\em Phys. Rev. B}, 91:155423, Apr 2015.

\bibitem{PhysRevB.46.4531}
J.-C. Charlier, J.-P. Michenaud, and X.~Gonze.
\newblock {F}irst-{P}rinciples {S}tudy of the {E}lectronic {P}roperties of
  {S}imple {H}exagonal {G}raphite.
\newblock {\em Phys. Rev. B}, 46:4531--4539, Aug 1992.

\bibitem{PhysRevLett.102.015501}
Zheng Liu, Kazu Suenaga, Peter J.~F. Harris, and Sumio Iijima.
\newblock {O}pen and {C}losed {E}dges of {G}raphene {L}ayers.
\newblock {\em Phys. Rev. Lett.}, 102:015501, Jan 2009.

\bibitem{PhysRevLett.109.206801}
A.~L. Rakhmanov, A.~V. Rozhkov, A.~O. Sboychakov, and Franco Nori.
\newblock {I}nstabilities of the $aa$-{S}tacked {G}raphene {B}ilayer.
\newblock {\em Phys. Rev. Lett.}, 109:206801, Nov 2012.

\bibitem{PhysRevB.90.155415}
R.~S. Akzyanov, A.~O. Sboychakov, A.~V. Rozhkov, A.~L. Rakhmanov, and Franco
  Nori.
\newblock $aa$-{S}tacked {B}ilayer {G}raphene in an {A}pplied {E}lectric
  {F}ield: {T}unable {A}ntiferromagnetism and {C}oexisting {E}xciton {O}rder
  {P}arameter.
\newblock {\em Phys. Rev. B}, 90:155415, Oct 2014.

\bibitem{kresse-prb-54-1996}
G.~Kresse and J.~Furthmuller.
\newblock {E}fficient iterative schemes for ab initio total-energy calculations
  using a plane-wave basis set.
\newblock {\em Phys. Rev. B"}, 54:11169, 1996.

\bibitem{kresse-prb-47-1993}
G.~Kresse and J.~Hafner.
\newblock \textit{{A}b initio} molecular dynamics for liquid metals.
\newblock {\em Phys. Rev. B}, 47:R558, 1993.

\bibitem{kresse-cms-6-1996}
G.~Kresse and J.~Furthmuller.
\newblock {E}fficiency of ab-initio total energy calculations for metals and
  semiconductors using a plane-wave basis sefficient iterative schemes for ab
  initio total-energy calculations using a plane-wave basis set.
\newblock {\em Compt. Mat. Sci.}, 6:15, 1996.

\bibitem{PhysRevB.23.5048}
J.~P. Perdew and Alex Zunger.
\newblock Self-interaction correction to density-functional approximations for
  many-electron systems.
\newblock {\em Phys. Rev. B}, 23:5048--5079, May 1981.

\bibitem{PhysRevB.59.1758}
G.~Kresse and D.~Joubert.
\newblock From ultrasoft pseudopotentials to the projector augmented-wave
  method.
\newblock {\em Phys. Rev. B}, 59:1758--1775, Jan 1999.

\bibitem{PhysRevB.50.17953}
P.~E. Bl\"ochl.
\newblock Projector augmented-wave method.
\newblock {\em Phys. Rev. B}, 50:17953--17979, Dec 1994.

\bibitem{PhysRevB.89.064305}
Lianhua He, Fang Liu, Geoffroy Hautier, Micael J.~T. Oliveira, Miguel A.~L.
  Marques, Fernando~D. Vila, J.~J. Rehr, G.-M. Rignanese, and Aihui Zhou.
\newblock {A}ccuracy of {G}eneralized {G}radient {A}pproximation {F}unctionals
  for {D}ensity-{F}unctional {P}erturbation {T}heory {C}alculations.
\newblock {\em Phys. Rev. B}, 89:064305, Feb 2014.

\bibitem{PhysRevB.60.11427}
Fabio Favot and Andrea Dal~Corso.
\newblock {P}honon {D}ispersions: {P}erformance of the {G}eneralized {G}radient
  {A}pproximation.
\newblock {\em Phys. Rev. B}, 60:11427--11431, Oct 1999.

\bibitem{PhysRevB.91.144107}
E.~Lora da~Silva, Jonathan~M. Skelton, Stephen~C. Parker, and Aron Walsh.
\newblock {P}hase {S}tability and {T}ransformations in the {H}alide
  {P}erovskite {C}s{S}n{I}$_{3}$.
\newblock {\em Phys. Rev. B}, 91:144107, Apr 2015.

\bibitem{monkhorst-prb-13-1976}
H.~J. Monkhorst and J.~D. Pack.
\newblock {S}pecial {P}oints for {B}rillouin-{Z}one {I}ntegrations.
\newblock {\em Phys. Rev. B}, 13:5188, 1976.

\bibitem{parlinski-prl-78-1997}
K.~Parlinski, Z.~Q. Li, and Y.~Kawazoe.
\newblock {F}irst-{P}rinciples {D}etermination of the {S}oft {M}ode in {C}ubic
  {Z}r{O}$_2$.
\newblock {\em Phys. Rev. Lett.}, 78:4063, 1997.

\bibitem{chaput-prb-84-2001}
L.~Chaput, A.~Togo, I.~Tanaka, and G.~Hug.
\newblock {P}honon-phonon interactions in transition metals.
\newblock {\em Phys. Rev. B}, 84:094302, 2001.

\bibitem{phonopy}
A~Togo and I~Tanaka.
\newblock First principles phonon calculations in materials science.
\newblock {\em Scr. Mater.}, 108:1, Nov 2015.

\bibitem{togo-prb-78-2008}
A.~Togo, F.~Oba, and I.~Tanaka.
\newblock {F}irst-principles calculations of the ferroelastic transition
  between rutile-type and {C}a{C}l$_2$-type {S}i{O}$_2$ at high pressures.
\newblock {\em Phys. Rev. B}, 78:134106, 2008.

\bibitem{skelton-prb-89-2014}
J.~M. Skelton, S.~C. Parker, A.~Togo, I.~Tanaka, and A.~Walsh.
\newblock {T}hermal physics of the lead chalcogenides {P}b{S}, {P}b{S}e, and
  {P}b{T}e from first principles.
\newblock {\em Phys. Rev. B}, 89:205203, 2014.

\bibitem{yu-cardona-springer-1996}
P.~Y. Yu and M.~Cardona.
\newblock {\em {F}undamentals of {S}emiconductors: {P}hysics and {M}aterials
  {P}roperties}.
\newblock Number pp. 104. 1996.

\bibitem{gajdos-prb-73-2006}
M.~Gajdo\v{s}, K.~Hummer, G.~Kresse, J.~Furthm\"{u}ller, and F.~Bechstedt.
\newblock {L}inear {O}ptical {P}roperties in the {PAW} {M}ethodology.
\newblock {\em Phys. Rev. B}, 73:045112, 2006.

\bibitem{0957-4484-21-6-065711}
Yuehua Xu, Xiaowei Li, and Jinming Dong.
\newblock {I}nfrared and {R}aman {S}pectra of {AA}-{S}tacking {B}ilayer
  {G}raphene.
\newblock {\em Nanotechnology}, 21(6):065711.

\bibitem{PhysRevB.39.12598}
M.~Hanfland, H.~Beister, and K.~Syassen.
\newblock {G}raphite {U}nder {P}ressure: {E}quation of {S}tate and
  {F}irst-{O}rder {R}aman {M}odes.
\newblock {\em Phys. Rev. B}, 39:12598--12603, Jun 1989.

\bibitem{doi:10.1063/1.2975333}
Jae-Kap Lee, Seung-Cheol Lee, Jae-Pyoung Ahn, Soo-Chul Kim, John I.~B. Wilson,
  and Phillip John.
\newblock {T}he {G}rowth of {AA} {G}raphite on (111) {D}iamond.
\newblock {\em The Journal of Chemical Physics}, 129(23):234709, 2008.

\bibitem{PhysRevLett.99.216802}
Eduardo~V. Castro, K.~S. Novoselov, S.~V. Morozov, N.~M.~R. Peres, J.~M.
  B.~Lopes dos Santos, Johan Nilsson, F.~Guinea, A.~K. Geim, and A.~H.~Castro
  Neto.
\newblock {B}iased {B}ilayer {G}raphene: {S}emiconductor with a {G}ap {T}unable
  by the {E}lectric {F}ield {E}ffect.
\newblock {\em Phys. Rev. Lett.}, 99:216802, Nov 2007.

\bibitem{sem}
G.~Semenoff.
\newblock {C}ondensed-{M}atter {S}imulation of a {T}hree-{D}imensional
  {A}nomaly.
\newblock {\em Phys. Rev. Lett.}, 53:2449, 1984.

\bibitem{PhysRevB.89.041405}
Brian Skinner, B.~I. Shklovskii, and M.~B. Voloshin.
\newblock {B}ound {S}tate {E}nergy of a {C}oulomb {I}mpurity in {G}apped
  {B}ilayer {G}raphene.
\newblock {\em Phys. Rev. B}, 89:041405, Jan 2014.

\bibitem{PhysRevB.74.161403}
Edward McCann.
\newblock {A}symmetry {G}ap in the {E}lectronic {B}and {S}tructure of {B}ilayer
  {G}raphene.
\newblock {\em Phys. Rev. B}, 74:161403, Oct 2006.

\bibitem{PhysRevB.76.035439}
M.~Mohr, J.~Maultzsch, E.~Dobard\ifmmode \check{z}\else
  \v{z}\fi{}i\ifmmode~\acute{c}\else \'{c}\fi{}, S.~Reich, I.~Milo\ifmmode
  \check{s}\else \v{s}\fi{}evi\ifmmode~\acute{c}\else \'{c}\fi{},
  M.~Damnjanovi\ifmmode~\acute{c}\else \'{c}\fi{}, A.~Bosak, M.~Krisch, and
  C.~Thomsen.
\newblock {P}honon {D}ispersion of {G}raphite by {I}nelastic {X}-{R}ay
  {S}cattering.
\newblock {\em Phys. Rev. B}, 76:035439, Jul 2007.

\bibitem{Cocemasov}
{P}honons in {B}ilayer {G}raphene.
\newblock \url{http://tkea.com.ua/siet/archive/2013-t2/130.pdf}.

\bibitem{ScientRepts.7.43956.2017}
Y.~Chen S.~Li.
\newblock {T}hermal {T}ransport and {A}nharmonic {P}honons in {S}trained
  {M}onolayer {H}exagonal {B}oron {N}itride.
\newblock {\em Scient. Rep.}, 7.

\bibitem{jiang-jpcom-27-2015}
S.~V. Morozov D. Jiang M. I. Katsnelson I. V. Grigorieva S. V. Dubonos A.
  A.~Firsov K.~S.~Novoselov, A. K.~Geim.
\newblock {A} {R}eview on the {F}lexural {M}ode of {G}raphene: {L}attice
  {D}ynamics, {T}hermal {C}onduction, {T}hermal {E}xpansion, {E}lasticity and
  {N}anomechanical {R}esonance.
\newblock {\em J. Phys.: Condens. Matter}, 27:083001, 2015.

\bibitem{PhysRevB.88.035428}
Alexandr~I. Cocemasov, Denis~L. Nika, and Alexander~A. Balandin.
\newblock {P}honons in {T}wisted {B}ilayer {G}raphene.
\newblock {\em Phys. Rev. B}, 88:035428, Jul 2013.

\bibitem{castro-graphite}
E.~V. Castro, N~.M.~R. Peres, J.~M. B.~Lopes dos Santos, F.~Guinea, and
  A.~H.~Castro Neto.
\newblock {\em Strongly Correlated Systems, Coherence and Entanglement}.
\newblock World Scientific Publishing Co. Pte. Ltd., 2007.

\bibitem{charlier_prb_1991}
J.~C. Charlier, X.~Gonze, and J.~P. Michenaud.
\newblock First-principles study of the electronic properties of graphite.
\newblock 43:4579, 1991.

\end{thebibliography}

\end{document}